\documentclass[12pt]{article}
\usepackage{amssymb}
\usepackage{amsmath}
\usepackage{amsthm}
\usepackage{bm} 
\usepackage{graphics}
\usepackage{graphicx}
\usepackage[T1]{fontenc}
\usepackage{longtable}
\usepackage{natbib}
\usepackage{setspace}
\usepackage{times}

\usepackage{epsfig}
\usepackage{fancyvrb}
\usepackage{multirow}
\usepackage[affil-it]{authblk}
\usepackage{verbatim}
\usepackage{tablefootnote}
\usepackage[explicit]{titlesec}
\usepackage{dcolumn}

\usepackage{subcaption}

\usepackage{tikz}
\usetikzlibrary{shapes,snakes, graphs}

\usetikzlibrary{arrows,%
                petri,%
                topaths}%
\usepackage{tkz-berge}
\usepackage{pgf}
\usepackage{tikz}
\usetikzlibrary{arrows,automata}
\usetikzlibrary{decorations.markings}

\newtheorem{theorem}{Theorem}
\newtheorem*{theorem*}{Theorem}
\newtheorem{lemma}{Lemma}

\newtheorem{Assumption}{Assumption}

\newcommand{\p}{\textrm{Pr}}

\def\ci{\mbox{\ensuremath{\perp\!\!\!\perp}}}

\def\logit{\textrm{logit}}
\def\expit{\textrm{expit}}

\DeclareMathOperator*{\argmin}{arg\,min}

\titleformat{\section}
  {\normalfont\Large\bfseries\centering}{\thesection.}{1em}{\MakeUppercase{#1}}
  
\begin{document}

\allowdisplaybreaks

\def\spacingset#1{\renewcommand{\baselinestretch}%
{#1}\small\normalsize} \spacingset{1}

\title{\bf Causal Inference When Counterfactuals Depend on the Proportion of All Subjects Exposed}
\author{
Caleb H. Miles$^{1,*}$, 
Maya Petersen$^{2,3,**}$, 
Mark J. van der Laan$^{2,4,***}$ \\
$^{1}$Department of Biostatistics, Columbia Mailman School of Public Health, New York, New York, U.S.A. \\
$^{2}$Division of Biostatistics, University of California at Berkeley, Berkeley, California, U.S.A. \\
$^{3}$Division of Epidemiology, University of California at Berkeley, Berkeley, California, U.S.A. \\
$^{4}$Department of Statistics, University of California at Berkeley, Berkeley, California, U.S.A. \hspace{.2cm}}
\date{}
\maketitle
\bigskip

\begin{abstract}
\noindent The assumption that no subject's exposure affects another subject's outcome, known as the no-interference assumption, has long held a foundational position in the study of causal inference. However, this assumption may be violated in many settings, and in recent years has been relaxed considerably. Often this has been achieved with either the aid of a known underlying network, or the assumption that the population can be partitioned into separate groups, between which there is no interference, and within which each subject's outcome may be affected by all the other subjects in the group via the proportion exposed (the stratified interference assumption). In this paper, we instead consider a complete interference setting, in which each subject affects every other subject's outcome. In particular, we make the stratified interference assumption for a single group consisting of the entire sample. This can occur when the exposure is a shared resource whose efficacy is modified by the number of subjects among whom it is shared. We show that a targeted maximum likelihood estimator for the i.i.d.~setting can be used to estimate a class of causal parameters that includes direct effects and overall effects under certain interventions. This estimator remains doubly-robust, semiparametric efficient, and continues to allow for incorporation of machine learning under our model. We conduct a simulation study, and present results from a data application where we study the effect of a nurse-based triage system on the outcomes of patients receiving HIV care in Kenyan health clinics.
\end{abstract}

\noindent%
\vfill

\newpage
\spacingset{1.45} 

\section{Introduction}
\label{s:intro}


Historically, the field of causal inference has relied on an assumption of no interference between subjects, meaning that one subject's exposure may not affect another subject's outcome \citep{cox1958planning}. This is one part of the stable unit treatment value assumption (SUTVA) \citep{rubin1980randomization}, which is invoked to ensure that counterfactuals (or potential outcomes) are well-defined. Counterfactuals are the hypothetical outcomes that would be observed had, possibly contrary to fact, a subject's exposure been set to a certain value. Under SUTVA, for a binary exposure there are two possible counterfactuals for each subject: the outcome we would see had they been exposed, and their outcome had they not been exposed. The fundamental problem of causal inference is that only one of these is observed for each subject, and a primary aim of causal inference is to borrow information from the unexposed to learn about exposed subjects' 
outcomes had they not been exposed, 
and vice versa. 

However, when subjects' exposures affect one another's outcomes, these two counterfactuals are not well defined, since the outcome we would observe had a subject been exposed may differ depending on other subjects' exposure levels. Thus, traditional causal estimands relying on these counterfactuals are not well defined, and there is ambiguity in what the causal quantity of interest should be in such settings. Further, without additional assumptions regarding the nature of the interference, 
there is no possibility for any subject of borrowing information across exposure levels to learn about counterfactuals under exposure levels that are not observed, since all subjects share the same observed exposure vector.



The reality is that SUTVA may often be violated in practice. In fact, in many cases, effects due to interference (often referred to as spillover effects) are themselves of interest, such as the herd immunity that can arise from vaccinations. Recently, a literature has emerged relaxing the assumption of no interference \citep[among others]{sobel2006randomized,rosenbaum2007interference,hudgens2008toward,
tchetgen2012causal,toulis2013estimation,basse2017analyzing}. Many methods rely on the assumption of \emph{partial interference} \citep{sobel2006randomized}, i.e., that subjects can be partitioned into groups such that a subject in one group may not interfere with a subject in another. These methods tend to rely on the number of distinct groups being relatively large, rather than relying on the sample sizes of the groups themselves being large (one exception is discussed in \cite{liu2014large}). One general approach for relaxing the partial interference assumption uses data on the underlying network connecting the subjects. The network is assumed to characterize the structure of the interference by specifying that only subjects who are connected in the network may interfere with one another. Methods using this approach has been proposed by \cite{toulis2013estimation,van2014causal,sofrygin2015semi,aronow2017estimating,
ogburn2017causal}.

In this paper, we consider an alternative relaxation of the partial interference assumption. Instead of assuming an explicit network structure through which interference operates, we allow for direct interference between all pairs of units, which we refer to as \emph{complete interference}. That is, we consider a trivial network that is completely connected. To make progress toward identification, we make the commonly-invoked \emph{stratified interference} assumption \citep{hudgens2008toward}, which states that each subject's outcome may be affected by all other subjects' exposures, but only via the total number (or equivalently, the proportion) of subjects who are exposed. This reduces the dimension of the 
function of all subjects' exposures upon which the counterfactuals are assumed to depend, from $n$ to two, recovering identifiability. This assumption is especially plausible when the interference is due to the strength of the exposure varying with how many subjects are receiving it, for instance when the exposure is a shared resource.

Examples of treatment effects varying with the proportion of subjects exposed are prevalent in the economics literature (see \cite{heckman1998general} and references in \cite{abbring2007econometric}). This literature overcomes the above identifiability challenge by leveraging stronger assumptions in the form of general equilibrium models, which 
model how treatment effects vary with the proportion of subjects exposed due to market mechanisms, permitting one to extrapolate effect estimates to settings with proportions of exposed subjects not observed. 
By contrast, our approach is agnostic toward this relationship. 

Our setting is closely related to one considered by \cite{liu2014large}. Namely, they consider one setting in which partial interference holds, and the number of groups is not large, but the number of subjects in each group is. They develop asymptotic results for certain interference causal parameters under the stratified interference assumption, including group-specific causal parameters, which corresponds to the single group setting we consider here. Two primary differences in our work are that (i) we consider an observational study setting, whereas they assume randomization at the group level, and (ii) our causal framework is model based, 
whereas they operate under the potential outcomes framework.

When only one group of completely connected units is observed, we find that only a certain class of effects is identifiable under our model without making stronger assumptions. This is a class of effects under interventions that preserve 
the overall proportion of subjects exposed, 
since the data only support one observation of the proportion exposed. This class contains analogous effects to the 
direct effects and overall effects 
of \cite{hudgens2008toward}.


We show that despite the dependence induced by interference, under stratified interference and standard causal assumptions, 
certain traditional i.i.d.~causal estimators can be used to estimate interference causal parameters. This is novel and important for two reasons: (i) it is a first solution for the problem of complete interference (i.e., a single block with 
a completely-connected network) in an observational study setting, and (ii) it is interesting and elucidating that a traditional estimator can be used for a non-traditional estimand. In particular, we focus on a targeted maximum likelihood estimator (TMLE), showing that it remains doubly robust, and permits the use of data-adaptive estimation of nuisance functions.


As a real data example, we examine the Low Risk Express Care (LREC) program implemented in Kenyan health clinics providing HIV care. Patients at clinics implementing the LREC program who were clinically stable (or ``low-risk") were eligible to have a subset of their clinical care tasks shifted from clinical officers to other health care professionals, such as nurses. 
Following program implementation, a non-randomized subset of eligible patients were enrolled in the task-shifting program. Previously, \cite{tran2016evaluating} analyzed data collected from this program under an assumption of no interference. However, they expressed concern that interference might in fact be present, as the effect of individual-level enrollment in the program is likely to differ depending on the number of other patients enrolled, and the corresponding workloads of 
health providers. We demonstrate properties of our estimator in finite samples via a simulation study, and apply our methodology to the LREC data to study the effect of the nurse-based task-shifting program on risk of death or loss to follow-up among patients receiving HIV care at a Kenyan clinic that implemented the LREC program.

\section{Defining the Model \& Target Parameters}
\label{s:defining}
\subsection{The Statistical and Causal Models}
To formalize our setting, let the observed data be $\bm{O^n}=\{(\bm{W_1},A_1,Y_1),\ldots,(\bm{W_n},A_n,Y_n)\}$, where $n$ is the sample size, and for subject $i$, $\bm{W_i}$ is a vector of baseline covariates, $A_i$ is a binary exposure indicator, and $Y_i$ is the outcome of interest, assumed to have bounded support. (Boundedness of the outcome will only be invoked for purposes of estimation and inference; for our identification results, the existence of its conditional expectation is sufficient.) 
Consider a model for $Y_i$ that only depends on $A_i$, $\bm{W_i}$, and $\bar{A}\equiv n^{-1}\sum_{i=1}^nA_i$, where we use the over-bar notation throughout to denote the sample mean. Thus, the dependence of each subject's outcome on $\bar{A}$ reflects the complete interference setting, since $\bar{A}$ involves every other subject in the sample's exposure. Let $q_{W,0}$ be the true probability density function of $\bm{W}$, 
$g_0$ be the true observed conditional exposure-assignment mechanism given $\bm{W_i}$ (which is known under a randomized controlled trial, but is otherwise unknown), and $q_{Y,0}$ be the true conditional probability density function of $Y$ given $A_i$, $\bm{W_i}$, and $\bar{A}$. 
Further, let $\bar{Q}_0\{\bm{W_i},A_i,k_n(\bar{A})\}\equiv E_{Y,0}\{Y_i\mid \bm{W_i},A_i,k_n(\bar{A})\}$, where each is a common function for all $i$ and $n$, and $k_n$ is some function that may depend on $n$.

The presence of the function $k_n$ is simply a relaxation of the outcome regression function, permitting dependence on $\bm{A^n}$ to vary with $n$ rather than simply being a fixed function of $\bar{A}$ with respect to $n$. For instance, rather than depending on the proportion exposed, one might suspect that the outcome regression depends on the total number of subjects exposed. In the former case, $k_n$ is simply the identity function; in the latter, $k_n(\bar{A})=n\bar{A}$. 
Note, however, that $\bar{Q}_0$ is naturally restricted by the assumption that $Y$ is bounded when $k_n(\bar{A})=n\bar{A}$; it cannot be linear in $k_n(\bar{A})$, for instance. 

It may often be the case that subjects' outcomes depend not only on the exposures of the other subjects sampled, but on the exposures of all subjects in some larger population from which they were sampled. In this case, $k_n$ can be thought of as a function mapping the proportion exposed in the sample to the proportion exposed in the larger population. We will only consider hypothetical interventions on the observed subjects, and so the exposures of subjects outside of the sample can be thought of as fixed. 
Our data example fits into this setting; in our case, it is more reasonable to model $\bar{Q}_0$ as a function of the proportion of all patients in the clinic who are task shifted, rather than of just the cohort of patients sampled (i.e., deemed eligible in the selected time window). As such, $k_n(\bar{A})$ can be defined as the proportion of all patients in the clinic task shifted.

When only one causally-connecting group is observed, knowledge of $k_n$ is in fact unnecessary for performing estimation, since, as we will see, we will be conditioning on $\bar{A}$, and hence on $k_n(\bar{A})$. However, specifying $k_n$ is useful for the purposes of interpretation, as the causal estimands we consider will be data adaptive with respect to $k_n(\bar{A})$, and will therefore only hold relevance to settings with a common value of $k_n(\bar{A})$. Also, its specification is required if, upon observing multiple groups with varying values of $k_n(\bar{A})$, one wishes to model how a causal parameter varies with $k_n(\bar{A})$ (see Web Appendix A for discussion of this setting).

We now discuss the LREC data and define the above notation in the context of this example. Patients were observed repeatedly over a period between 2006 and 2009. We discretize these observations into 90-day intervals (corresponding to the approximate frequency of clinic visits), and for each clinic select as baseline the discretized time period with the most observed exposed patients (or unexposed, if fewer). Patients were included in the analysis if they had at least one visit to the clinic during the baseline time period, were eligible to be task-shifted at their visit, and were not previously enrolled in the task-shifting program. The binary individual exposure indicator $A_i$ is defined as an indicator of enrollment into the task-shifting program at the baseline visit (irrespective of enrollment at subsequent time points). 
The outcome $Y_i$ is defined as an indicator that the patient died or was lost to follow up within 270 days of baseline. As noted in \cite{tran2016evaluating}, ``Patients who do not return for continuing HIV care are subject to higher risk of complications and health decline, 
placing them at unnecessarily higher mortality rates.'' Lastly, $\bm{W_i}$ consists of the static baseline covariates age, sex, adherence to antiretroviral therapy (ART), CD4 count, indicator of protease inhibitor-based ART regimen, WHO disease stage, and indicator of tuberculosis treatment at beginning of ART. 


In our example, the stratified interference assumption implies that a patient's retention and mortality is potentially influenced by his or her own enrollment in the program as well as by the proportion of all HIV patients concurrently enrolled, but not by the specific allocation of the exposure to the other patients. This assumption is more plausible if all patients require an equal intensity of care -- an assumption that, while unlikely to hold completely, might be a reasonable approximation, since by definition all patients in the analysis are clinically stable enough to meet eligibility criteria for task shifting.

Let $\mathcal{M}_{si}$ be the nonparametric model under stratified interference that leaves $(q_{Y},g,q_W)$ unrestricted (apart from support conditions). We assume that the observed data follow the distribution
$p_0 \allowbreak (O^n) \allowbreak = \allowbreak \prod_{i=1}^n \allowbreak q_{Y,0}\{ \allowbreak Y_i \allowbreak \mid  \allowbreak \bm{W_i}, \allowbreak A_i, \allowbreak k_n( \allowbreak \bar{A})\} \allowbreak g_0( \allowbreak A_i \allowbreak \mid  \allowbreak \bm{W_i}) \allowbreak q_{W,0}( \allowbreak \bm{W_i})\in\mathcal{M}_{si}$. That is, we assume that patient $i$'s outcome depends on the exposure assignment vector $\bm{A^n}\equiv(A_1,\ldots,A_n)$ and covariate matrix $\bm{W^n}\equiv(\bm{W_1},\ldots,\bm{W_n})$ only through $A_i$, $\bm{W_i}$, and $k_n(\bar{A})$. 

Next, we define our causal model, which involves additional structural assumptions. We posit a nonparametric structural equation model (NPSEM), which states that each observed variable is generated by a distinct, arbitrary function, denoted $f$, of (i) other observed variables, and (ii) a distinct random disturbance, denoted $U$, where we use subscripts on $f$ and $U$ to identify these with their corresponding observed variable. Specifically, we posit 
\begin{align*}
\bm{W_i}&= \bm{f_W}(\bm{U^W_i})\\
A_i&=f_A(\bm{W_i},U^A_i)\tag{1}\\
Y_i&=f_Y\{\bm{W_i},A_i,k_n(\bar{A}),U^Y_i\},
\end{align*}
where the random disturbances $\bm{U^W_i}$, $U^A_i$, and $U^Y_i$ are assumed to all be mutually independent both within and across $i$. This mutual independence assumption is structural, and implies the generative nature of the model. That is, this implies that a hypothetical intervention to set a variable to a certain value can impact all variables that are functions of the intervened variable in the NPSEM. 
The mutual independence assumption also implies that the only source of dependence between subjects is due to their outcomes' common dependence on $\bar{A}$. This assumption can be violated if there is residual confounding, or if there are factors beyond $\bar{A}$ inducing dependence between subjects.


This model is a straightforward extension of the NPSEM for a point treatment setting with no interference; the only difference is the presence of $k_n(\bar{A})$ in $f_Y$. 
Also, it is an extreme case of stratified interference under the partial interference assumption, which states that $\{1,\ldots,n\}$ can be partitioned into $J$ groups $\{k_1,\ldots,k_J\}$ 
such that for each $i$, $Y_{i}=f_Y\{\bm{W_{i}},A_{i},k_n(\bar{A}_j),U^Y_i\}$ for $i\in k_j$, where $\bar{A}_j$ is the sample mean of $A$ among $k_j$; in our case $J=1$.

\subsection{Interventions, counterfactuals, and causal estimands}
We now consider hypothetical interventions and their implications under the NPSEM. These do not correspond to the true exposure assignment distribution, 
but are rather interventions under which we might be interested in knowing what happens with subjects' outcomes. Consider a hypothetical intervention setting $\bm{A^n}$ to a particular value $\bm{a^n}$. The NPSEM implies that the system of equations (1) yields $\bm{W_i}=\bm{f_W}(\bm{U_i^W})$, $A_i=a_i$, $Y_i^{a_i,\allowbreak\bar{a}}\allowbreak =\allowbreak f_Y\{\allowbreak \bm{W_i}, \allowbreak a_i, \allowbreak k_n(\bar{a}),\allowbreak U_i^Y\}$, where $Y_i^{a_i,\bar{a}}$ is the counterfactual we would observe for subject $i$ had they been assigned exposure level $a_i$ and had $\bar{a}$ of subjects in the sample been exposed. Thus, the model implies that $Y_i^{a_i,\bar{a}}\ci A_i\mid \bm{W_i}$ for each $i$ and all levels of $\bm{a^n}$, i.e., the observed covariates are sufficient to control for confounding. The NPSEM also encodes the consistency assumption: when $\bm{A^n}=\bm{a^n}$, $Y_i^{a_i,\bar{a}}=f_Y\{\bm{W_i},a_i,k_n(\bar{a}),U^Y_i\}=f_Y\{\bm{W_i},A_i,k_n(\bar{A}),U^Y_i\}=Y_i$ for each $i$, and for all levels of $\bm{a^n}$. These latter two implications will be useful for identification. 

We will consider two types of counterfactuals. The first corresponds to the previous example and is denoted by a double superscript. The first argument in the superscript indicates subject $i$'s exposure level; the second indicates the proportion exposed in the sample. Under the stratified interference assumption, this counterfactual is the same for subject $i$ for all exposure vectors that generate the same $a_i$ and $\bar{a}$, so we need only conceive of a hypothetical intervention setting $A_i$ to some level $a$ and the proportion $\bar{A}$ to some level $\pi$ to define the counterfactual under this intervention, viz., $Y_i^{a,\pi}$. Defining such a counterfactual for all $n$ subjects, we then define their sample mean to be $\bar{Y}^{a,\pi}\equiv n^{-1}\sum_{i=1}^nY_i^{a,\pi}$. Our first class of target causal parameters is the conditional mean of this sample average given the observed covariate values: $E\left(\bar{Y}^{a,\pi}\mid \bm{W^n}\right)$. These parameters correspond to effects in populations of subjects with the same covariate values as those in our current sample, therefore inference for these parameters will not account for variability induced by the sampling of these covariates. Such parameters are naturally relevant to the group of subjects actually sampled, which we argue will often be the population of greatest interest for inference. We consider parameters that marginalize over the distribution of the baseline covariates in Web Appendix D.

For a given proportion of exposed subjects, $\pi$, we can define a direct effect as a contrast of two of the above parameters: $E\left(\bar{Y}^{1,\pi}\mid \bm{W^n}\right)-E\left(\bar{Y}^{0,\pi}\mid \bm{W^n}\right)$. This contrasts two parameters for which the proportion exposed is held fixed at $\pi$, and the individual exposure level is changed from unexposed to exposed. In our LREC example, this corresponds to the direct effect of an individual's enrollment in the task-shifting program, holding the proportion of enrolled subjects constant at $\pi$. 
Relatedly, \cite{hudgens2008toward} define a \emph{group average direct causal effect} under complete randomization of a single group, which depends on the experimental exposure assignment mechanism. However, under stratified interference, this estimand is invariant to this assignment mechanism apart from the coverage level $\pi$, and using our notation, is equal to $\bar{Y}^{0,\pi}-\bar{Y}^{1,\pi}$. They operate under the potential outcomes framework, in which counterfactuals are considered to be fixed. The only differences in our estimand are that we consider these to be random, and take the conditional expectation given covariates, and more superficially, we flip the order of intervention level.





More generally, we can consider counterfactuals arising from hypothetical interventions that are potentially dynamic and/or stochastic, rather than simply setting the exposure to the same fixed value for all subjects. A stochastic intervention assigns the exposure with some non-degenerate probability. For example, in the LREC example, an intervention that randomly shifts the care tasks of 50\% of patients would be a stochastic intervention. A dynamic intervention is an intervention that depends on covariates. For example, in the LREC example, an intervention that shifts care tasks for the 50\% of patients with the highest CD4 counts would be a dynamic intervention.

Consider a general hypothetical intervention $g^*$ that is a joint distribution that can be dependent across subjects. We denote an exposure vector generated under the intervention $g^*$ by $\bm{A^{*n}}$ such that $\bm{A^{*n}}\mid \bm{W^n}\sim g^*$. Throughout, we will only consider row-exchangeable interventions, meaning that 
the order in which subjects are listed in the data frame is irrelevant to their exposure probability \citep{imbens2015causal}. For such an intervention, we define the counterfactual $Y_i^{g^*\!,\pi}\equiv f_Y\{\bm{W_i},A^*_i,k_n(\pi),U^Y_i\}$, which is the outcome we would have seen had the exposure been assigned according to $g^*$, and had the overall proportion of subjects exposed been $\pi$. This intervention need not place any restriction on $\bar{A}^*$; the counterfactuals $Y_i^{g^*\!,\pi}$ are still well defined even when $\bar{A}^*\neq\pi$. In fact, this class of counterfactuals contains the two contrasted in our above definition of the direct effect, where the interventions being compared are $g^*(\bm{A^n}=\bm{1}\mid \bm{W^n})=1$ and $g^*(\bm{A^n}=\bm{0}\mid \bm{W^n})=1$. We define the analogous causal parameter to $\psi^{a\!,\pi}$ that replaces the intervention setting $A_i$ to $a$ for all $i$ with the more general intervention $g^*$ as $\psi^{g^*\!,\pi}\equiv E(\bar{Y}^{g^*\!,\pi}\mid W^n)$. 

The second type of counterfactual we consider is denoted with a single superscript, which indicates the hypothetical intervention distribution for the entire exposure vector. Intervening on $\bm{A^n}$ according to $g^*$, the NPSEM implies that the system of equations (1) yields, for all $i$, $\bm{W_i}=\bm{f_W}(\bm{U_i^W})$, $\bm{A^{*n}}\sim g^*(\bm{a^n}\mid \bm{W^n})$, $Y_i^{g^*}=f_Y\{\bm{W_i}, A^*_i, k_n(\bar{A}^*),U_i^Y\}$, where $Y_i^{g^*}$ is the counterfactual we would observe for subject $i$ had the entire exposure vector been assigned according to $g^*$. We term these \emph{overall effect counterfactuals}. We define the sample average of these counterfactuals for all subjects under a hypothetical intervention $g^*$ as $\bar{Y}^{g^*}\equiv n^{-1}\sum_{i=1}^nY_i^{g^*}$, and the corresponding overall effect parameter as $\psi^{g^*}\equiv E\left(\bar{Y}^{g^*}\mid \bm{W^n}\right)$. 
For any two interventions $g^*$ and $g^{\dag}$, we refer to the contrast $\psi^{g^*}-\psi^{g^{\dag}}$ as an overall effect.

Overall effect parameters and overall effects are superpopulation analogs to the \emph{marginal group average potential outcomes} and \emph{group average overall causal effects}, respectively, defined by \cite{hudgens2008toward}. Once again, \cite{hudgens2008toward} operate under a potential outcomes framework, so that the only component of $Y_i^{g^*}\allowbreak =\allowbreak f_Y\{\allowbreak \bm{W_i}, \allowbreak A_i^*, \allowbreak k_n(\bar{A}^*), \allowbreak U_i^Y\}$ 
they consider to be random is $\bm{A^{*n}}$. Thus, they define the marginal group average potential outcome to be the expectation of $\bar{Y}^{g^*}$ only with respect to $\bm{A^{*n}}$, and implicitly condition on $\bm{W^n},\bm{U^{Y,n}}$. They define contrasts of any two such interventions to be group average overall causal effects. They focus on an experimental setting under complete randomization, such that $g^*$ corresponds to the observed experimental exposure assignment distribution. \cite{van2014causal} extends these parameters to observational study settings, and potentially dynamic/stochastic interventions that do not correspond to the observed exposure distribution. Our definition of overall effect parameters are equivalent to the causal estimands of \cite{van2014causal} that condition on the observed covariates.

\section{Nonparametric Identification}
Having defined the above causal parameters, we now wish to nonparametrically identify them with a consistently estimable quantity using observed data. To do so, we first distinguish between two senses of the term identification. 
We say a parameter is \emph{causally nonparametrically identifiable} if it can be expressed as a functional of the observed data distribution under a nonparametric model. We say it is \emph{statistically nonparametrically identifiable} if it can be consistently estimated (however slowly) from observed data under a nonparametric model. The former does not necessarily imply the latter, as we will see.

We begin with causal nonparametric identification. First, we fix some new notation for convenience. Due to the row-exchangeability of $g^*$, $\sum_{\bm{a^n_{-i}}}g^*(\bm{a^n}\mid \bm{W^n})$ depends on $i$ only through $\bm{W_i}$ for each $i\in\{1,\ldots,n\}$. Define $\theta_n\equiv\{\bm{W^n},k_n(\bar{A})\}$, and $g_{\theta_n}^*(a\mid \bm{w})$ to be a function that is equal to $\sum_{\bm{a^n_{-i}}}g^*(\bm{a^n}\mid \bm{W^n})$ at $\bm{w}=\bm{W_i}$ for each $i\in\{1,\ldots,n\}$. The (potential) dependence of $g^*_{\theta_n}$ on $\bm{W^n_{-i}}$ is acknowledged notationally in the subscript $\theta_n$. This is a known function, as $g^*$ is user-specified. 
For identification of $\psi^{g^*\!,\pi}$ and $\psi^{g^*}$, respectively, we need the following positivity assumptions to hold:
\begin{Assumption}
$g^*_{\theta_n}(a\mid \bm{W_i})/\mathrm{Pr}_{p_0}\{A_i=a\mid \bm{W_i}, \bar{A}=\pi\}<\infty$ for all $i$ and $a\in\{0,1\}$.
\end{Assumption}
\begin{Assumption}
$g^*(\bm{a^n}\mid \bm{W^n})/\mathrm{Pr}_{p_0}(A_i=a, \bar{A}=\bar{a}\mid \bm{W_i})<\infty$ for all $i$ and $\bm{a^n}\in\{0,1\}^n$.
\end{Assumption}
We have the following causal identification result for these parameters.
\begin{theorem}
Suppose the NPSEM defined in (1) contains the true set of underlying counterfactual distributions. 
Under Assumptions 1 and 2, respectively, we have,
\begin{align*}
\psi^{g^*\!,\pi} &= \frac{1}{n}\sum_{i=1}^n\sum_{a=0}^1\bar{Q}_{0}\{\bm{W_i},a,k_n(\pi)\}g_{\theta_n}^*(a\mid \bm{W_i})\\
\psi^{g^*} &= \frac{1}{n}\sum_{i=1}^n\sum_{\bm{a^n}}\bar{Q}_{0}\{\bm{W_i},a_i,k_n(\bar{a})\}g^*(\bm{a^n}\mid \bm{W^n}).
\end{align*}
\end{theorem}
All proofs are provided in Web Appendix G. 

While Theorem 1 gives an expression for $\psi^{g^*\!,\pi}$ as a functional of the observed data distribution, whether its identifying functional can be estimated consistently depends on the value of $\pi$. In order to estimate this quantity, it is necessary to have statistical support at $\bar{A}=\pi$. 
For instance, one could estimate $\psi^{g^*\!,\pi}$ by plugging in an estimate of the outcome regression function $\bar{Q}_{0}\{\bm{w},a,k_n(\pi)\}$ specifically at $\pi$, which requires observing $Y_i$ and $\bar{A}=\pi$ simultaneously for some subjects. However, we only observe one value of $\bar{A}$ for all subjects, and so for a fixed $\pi$, we will rarely observe outcomes when $\bar{A}=\pi$, and will do so with decreasing probability as sample size grows. For a fixed $\pi$, this outcome regression 
cannot be consistently estimated. Similarly, statistical nonparametric identification for $\psi^{g^*}$ relies on the choice of the hypothetical intervention $g^*$. If $g^*$ produces exposure vectors with a sample mean that do not correspond to the observed $\bar{A}$, then we encounter the same problem as for $\psi^{g^*\!,\pi}$. 


As a result, we will only be able to identify data-adaptive target parameters depending on $\bar{A}$ from our data without making strong assumptions. 
Specifically, we will only be able to perform inference on $\psi^{g^*\!,\pi} $ when $\pi = \bar{A}$. Substituting for $\pi$, we express counterfactuals under the intervention setting the individual-level exposure by $g^*$ and setting the proportion exposed to the observed proportion of subjects exposed as $Y_i^{g^*\!,\bar{A}}$, and the corresponding causal parameter as $\psi^{g^*\!,\bar{A}}$.

This causal parameter is in fact equivalent to the overall effect parameter $\psi^{g^*}$ when $g^*$ assigns $\bar{A}$ of subjects to exposure with probability one. 
We refer to such interventions as \emph{exposure reallocation schemes} (ERS), as they define schemes for reallocating the same number of exposures among the sample. 
One example of an ERS is complete randomization with probability $\bar{A}$, in which the support for $g^*$ is $\{\bm{a^{*n}}:\bar{a}^*=\bar{A}\}$, and each exposure-assignment vector has equal probability, such that each subject is equally likely to be exposed. In the LREC example, this would correspond to randomly shifting the care tasks of $\bar{A}$ of the patients. Another example is an intervention that takes a scalar function of each subject's covariates, and assigns the exposure to the $\sum_{i=1}^nA_i$ subjects with the highest values of this function. Such a dynamic (or personalized) reallocation scheme makes it possible to define causal parameters that, for example, quantify how outcomes could be improved by reallocating a resource-constrained intervention to the patients most likely to benefit from it. In the LREC example, if this function of a patient's covariates returned their CD4 count, this intervention would correspond to shifting the care tasks of the $\bar{A}$ with the highest CD4 counts, which could be viewed as a proxy for the sickest $\bar{A}$ patients. ERSs are necessarily dependent across subjects, as one subject's exposure will be determined given the other $n-1$ subjects' exposures. The exposure reallocation criterion can in fact be relaxed to allow for interventions that are independent across subjects, which we discuss in Web Appendix B.

We have the following statistical identification results 
for $\psi^{g^*\!,\bar{A}}$ and, under an ERS, $\psi^{g^*}$:
\begin{theorem}
Suppose the NPSEM defined in (1) contains the true set of underlying counterfactual distributions, and Assumption 1 holds for $\pi = \bar{A}$. Then 
\begin{align*}
\psi^{g^*\!,\bar{A}}=\frac{1}{n}\sum_{i=1}^n\sum_{a=0}^1\bar{Q}_{0}\{\bm{W_i},a,k_n(\bar{A})\}g_{\theta_n}^*(a\mid \bm{W_i})\equiv\Psi^{g^*\!,\bar{A}}(\bar{Q}_0).
\end{align*}
If $g^*$ is an ERS, $\psi^{g^*}=\psi^{g^*\!,\bar{A}}$, and is statistically nonparametrically identified by $\Psi^{g^*\!,\bar{A}}(\bar{Q}_0)$.
\end{theorem}
As these identifiable parameters are necessarily data adaptive, they will naturally vary with $\bar{A}$. Just how much they vary will depend on the variability of $\bar{A}$ itself 
and the sensitivity of $\bar{Q}_0$ to $\bar{A}$, which 
will be setting-specific. Thus, how interesting these parameters are is subjective, and will vary from setting to setting. However, we emphasize the point here that unless one is willing to make stronger extrapolating assumptions, these data-adaptive parameters are the only parameters we can hope to estimate consistently with data from a single group.

Spillover effects are another form of causal effects that may commonly be of interest in the presence of interference. However, 
these will generally be inestimable from our data when comparing interventions that result in different proportions of subjects assigned to exposure. 
We consider their estimation when many groups are observed, as well as the role of $k_n(\bar{A})$ as an effect modifier 
in Web Appendix A.

\section{Estimation \& Inference}
\label{s:estimation}
We now show that the semiparametric efficient estimator one would use under an assumption of no interference for a conditional average treatment effect (CATE) parameter under a potentially dynamic and/or stochastic intervention $g^*$ will also be consistent and asymptotically normal for $\Psi^{g^*\!,\bar{A}}(\bar{Q}_0)$ under $\mathcal{M}_{si}$ given an appropriate choice of $g^*_{\theta_n}$. 
We define the i.i.d.~causal and statistical models as the submodels of the NPSEM defined in (1) and $\mathcal{M}_{si}$, respectively, induced by imposing the no-interference assumption, i.e., by dropping the dependence of $Y$s on $\bar{A}$. 
The NPSEM defined in (1) reduces to $\bm{W_i}= \bm{\tilde{f}_W}(\bm{U^W_i})$, $A_i=\tilde{f}_A(\bm{W_i},U^A_i)$, $Y_i=\tilde{f}_Y(\bm{W_i},A_i,U^Y_i)$, and the statistical model $\mathcal{M}_{iid}$ becomes $\tilde{p}_{0}(\bm{O^n})=\prod_{i=1}^n\tilde{q}_{Y,0}(Y_i\mid A_i,\bm{W_i})\tilde{g}_0(A_i\mid \bm{W_i})\tilde{q}_{W,0}(\bm{W_i})$. The conditional-average causal target parameter is $\psi_{iid}^{g^*} \equiv n^{-1}\sum_{i=1}^nE(Y^{g^*}\mid \bm{W_i})$, where $Y^{g^*}\equiv \tilde{f}_Y(\bm{W},A^*,U^Y)$ and $A^*$ is the individual-level exposure assigned under intervention $g^*$. The parameter $\psi^{g^*}_{iid}$ is identified by $\Psi_{iid}^{g^*}(\bar{Q}_0)\equiv n^{-1}\sum_{i=1}^n\sum_{a=0}^1\bar{Q}_0(a,\bm{W_i})g^*(a\mid \bm{W_i})$, and the CATE is defined as the contrast in $\psi_{iid}^{g^*}$ comparing the individual-level interventions $g^*(1\mid \bm{W})=1$ and $g^*(1\mid \bm{W})=0$.

For a given semiparametric model and target estimand that is a functional of the observed data distribution, the TMLE is a substitution estimator that is semiparametric efficient. We review the TMLE for $\Psi_{iid}^{g^*}(\bar{Q}_0)$ in the i.i.d.~setting in Web Appendix C. 
To build intuition for why this standard estimator can be used to estimate the nonstandard interference parameter $\Psi^{g^*\!,\bar{A}}(\bar{Q}_0)$, we draw a correspondence between conditional distributions under $\mathcal{M}_{si}$ given $\theta_n=\{\bm{W^n},k_n(\bar{A})\}$ and conditional distributions given $\bm{W^n}$ under $\mathcal{M}_{iid}$. First observe that because the same $\bar{A}$ is observed for all subjects, any estimator will be conditioning on this common value by default. Intuitively, estimators that are consistent for $\bar{Q}_0$ and $g_0$ under $\mathcal{M}_{iid}$ will also be consistent for the functions $\bar{Q}_{\theta_n,0}\equiv\bar{Q}_0(\cdot,\theta_n)$ and $g_{\theta_n,0}\equiv g_0(\cdot\mid\cdot,\theta_n)$ 
that further condition on $\bar{A}$ under $\mathcal{M}_{si}$, since $\bar{A}$ is in fact observed and the same for each subject. 

Since our estimand is also conditional on $\bm{W^n}$, we are effectively operating under the conditional distribution given $\theta_n=\{\bm{W^n},k_n(\bar{A})\}$, which we define as $p_{\theta_n}(\bm{y^n},\bm{a^n}\mid \bm{w^n}, \bar{a})\equiv \left[\prod_{i=1}^nq_{Y}\{y_i\mid \bm{w_i}, a_i, k_n(\bar{a})\}\right]g^n(\bm{a^n}\mid \theta_n)$, where $g^n(\bm{a^n}\mid \theta_n)\equiv Pr_{p_0}(\bm{A^n}=\bm{a^n}\mid \{\bm{W^n},k_n(\bar{A})\}=\theta_n)$, which is the joint propensity score of the entire exposure vector conditional on $\bar{A}$ in addition to covariates. By contrast, a conditional distribution given $\bm{W^n}$ under the i.i.d.~model is $\tilde{p}(\bm{y^n},\bm{a^n}\mid \bm{w^n})=\prod_{i=1}^n\tilde{q}_{Y}(y_i\mid a_i,\bm{w_i})\tilde{g}(a_i\mid \bm{w_i})$. Consider a particular distribution in the latter model such that for $\theta_n$ sampled from the marginal stratified interference model $p$, $\tilde{p}$ satisfies $\tilde{q}_{Y}(y_i\mid \bm{w_i}, a_i)=q_{Y}\{y_i\mid \bm{w_i}, a_i, k_n(\bar{a})\}$. Then $p_{\theta_n}$ and $\tilde{p}$ differ only in their distribution of $\bm{A^n}$. Additionally, consider the conditional-average causal parameter $\Psi_{iid}^{g^*}(\bar{Q}_0)$ for an individual-level intervention $g^*$ equivalent to the intervention $g^*_{\theta_n}$ based on $\theta_n$ sampled from the interference model $p$. Then under the i.i.d.~model, the score corresponding to the $\tilde{g}$ component of the likelihood is in the nuisance tangent space of $\Psi_{iid}^{g^*}(Q_0)$, implying that the asymptotic distribution of any semiparametric efficient estimator of $\Psi_{iid}^{g^*}(Q_0)$ is invariant to the conditional distribution of $\bm{A^n}$ given $\bm{W^n}$. This means that under any distribution with an equivalent $q_Y$ component, the estimator that is semiparametric efficient with respect to the i.i.d.~model will have the same asymptotic distribution as it does under $\tilde{p}$. More specifically, since $p_{\theta_n}$ satisfies this criterion, and $\Psi^{g^*\!,\bar{A}}(\bar{Q}_0)$ corresponds to $\Psi_{iid}^{g^*}(\bar{Q}_0)$, this implies that the standard i.i.d.~semiparametric efficient estimator for $\Psi_{iid}^{g^*}(\bar{Q}_0)$ under the intervention $g^*_{\theta_n}$ has the same asymptotic distribution in $\mathcal{M}_{si}$ when used to estimate $\Psi^{g^*\!,\bar{A}}(\bar{Q}_0)$.

Put another way, conditional on $\theta_n$, a semiparametric efficient estimator will be asymptotically linear under $\mathcal{M}_{si}$, but its influence curve will be dependent across evaluations at different subjects' observations. However, since the $Y$s are conditionally independent and mean zero given $\{\bm{W^n},\bm{A^n}\}$, we can apply the Lindeberg central limit theorem for a sequence of independent, non-identically distributed random variables given $\{\bm{W^n},\bm{A^n}\}$. Therefore, the proposed estimation strategy is to replace the $g^*$ in the estimand in the standard TMLE with $g^*_{\theta_n}$, which remains known, and then proceed with the standard analysis ignoring the presence of interference. The following theorem formalizes its asymptotic properties.

\begin{theorem}
Suppose the technical 
conditions listed in Web Appendix F hold. Then for $\Psi^{g^*\!,\bar{A}}(\bar{Q}_0)$, the TMLE $\psi_n=\Psi_{iid}^{g^*_{\theta_n}}(\bar{Q}^*_n)$ for $\Psi_{iid}^{g^*_{\theta_n}}(\bar{Q}_0)$ under $\mathcal{M}_{iid}$ is consistent and asymptotically normal under $\mathcal{M}_{si}$, with asymptotic variance
\[\sigma^2_Y \equiv\lim\limits_{n\rightarrow\infty}n^{-1}\sum_{i=1}^n\left[\frac{g_{\theta_n}^*(A_i\mid \bm{W_i})}{g_{\theta_n,0}(A_i\mid \bm{W_i})}\left\{Y-\bar{Q}_{\theta_n,0}(A_i,\bm{W_i})\right\}\right]^2.\]
\end{theorem}
One can then use the variance estimator one would use for the TMLE as an estimator of $\Psi_{iid}^{g^*_{\theta_n}}(\bar{Q}_0)$ under $\mathcal{M}_{iid}$, and inference based on the model with no interference will be valid. 
One can use Wald tests and Wald-type confidence intervals just as in the setting with no interference. The augmented-inverse probability weighted (A-IPW) estimator for $\Psi_{iid}^{g^*_{\theta_n}}(\bar{Q}_0)$ (defined in Web Appendix C) has the same asymptotic properties, and its proof follows analogously to that of Theorem 3. A consequence of Theorem 3 is that if one is to naively estimate the CATE with a TMLE or A-IPW estimator ignoring the presence of interference, the resulting estimate can be interpreted as the direct effect of exposure in a setting in which $\bar{A}$ of the sample is exposed. Other estimators may have the same interpretation, though their asymptotic distributions 
appear to be more challenging to derive.

\section{Simulation Study}
\label{s:sims}
We now discuss results from a simulation study in order to demonstrate the finite-sample performance of our method. A more detailed review of the study is available in Web Appendix E. Our simulation design combines different sample sizes (50, 500, and 5000) and interference levels governed by the parameter $\beta$ (0,1,
10). 
The setting with $\beta=0$ corresponds to there being no interference. The setting with $\beta=1$ corresponds to a setting in which shifting everyone would completely negate the effect of the exposure. When $\beta>1$, the direction of the individual exposure effect reverses when enough individuals are exposed. 

We consider two data-adaptive causal parameters depending on the observed $\bar{A}$: (i) the direct effect when $\bar{A}$ of subjects are exposed, and (ii) the overall effect under the ERS that assigns subjects with the $S_n\equiv\sum_{i=1}^nA_i$ highest (scalar) $W$ values to exposure. 
For this data generating process, the intervention assigning the exposure to subjects with the highest values of $W$ corresponds with the optimal exposure reallocation scheme (OERS) within the class of all ERSs. 
We consider three different parametric models for estimation: one in which the outcome regression is correctly specified, but the propensity score is not, one in which the reverse is true, and one in which both are correctly specified. 

\begin{table}
\begin{center}
\caption{Simulation results for the direct effect and overall effect under the optimal exposure reallocation scheme (OERS) based on 5000 samples. MSE is mean squared error, and CP is the Monte Carlo coverage probability of the 95\% confidence interval.}
\label{t:one}
\centering
\begin{tabular}{c c l c r@{.}l r@{.}l r@{.}l c r@{.}l r@{.}l r@{.}l}
\\
\hline\hline
&&&&	\multicolumn{13}{c}{Target parameter}	\\
\cline{5-17}
&&&&	\multicolumn{6}{c}{Direct effect}	&&	\multicolumn{6}{c}{Overall effect -- OERS}\\
\cline{5-10} \cline{12-17}
$\bar{Q}$	&	$g$	&	$n$	&	$\beta$ &\multicolumn{2}{c}{Bias}&\multicolumn{2}{c}{MSE}&\multicolumn{2}{c}{CP}&&\multicolumn{2}{c}{Bias}&\multicolumn{2}{c}{MSE}&\multicolumn{2}{c}{CP}\\
\hline
Correctly	&	Correctly	&	50	&	0		&	0&010	&	0&10	&	0&916	&&	2&2e-4	&	0&027	&	0&948	\\
specified	&	specified	&	&	1		&	4&6e-3	&	0&10	&	0&917	&&	8&4e-5	&	0&027	&	0&951	\\
	&		&	&	10	&	-0&17	&	0&16	&	0&837	&&	0&026		&	0&028	&	0&942	\\
\cline{3-17}
	&		&	500	&	0		&	2&0e-3	&	9&7e-3	&	0&955	&&	3&0e-4	&	2&7e-3	&	0&960	\\
	&		&	&	1		&	-1&8e-3	&	9&6e-3	&	0&958	&&	1&4e-3	&	2&7e-3	&	0&962	\\
	&		&	&	10	&	-0&026	&	0&011	&	0&942	&&	5&7e-3	&	2&7e-3	&	0&960	\\
\cline{3-17}
	&		&	5000	&	0		&	1&1e-4	&	9&7e-4	&	0&958	&&	-1&5e-4	&	2&6e-4	&	0&964	\\
	&		&	&	1		&	8&6e-5	&	9&6e-4	&	0&961	&&	6&9e-5		&	2&6e-4	&	0&965	\\
	&		&	&	10	&	-3&0e-3	&	9&8e-4	&	0&961	&&	6&3e-4		&	2&7e-4	&	0&960	\\
\hline
Mis-	&	Correctly	&	50	&	0		&	8&7e-3	&	0&10	&	0&884	&&	4&5e-3	&	0&032	&	0&940	\\
specified	&	specified	&			&	1		&	8&0e-4	&	0&10	&	0&885	&&	2&5e-3	&	0&031	&	0&938	\\
	&		&			&	10	&	-0&16	&	0&15	&	0&806	&&	5&0e-3	&	0&048	&	0&983	\\
\cline{3-17}
	&		&	500	&	0	&	-2&4e-4	&	9&8e-3	&	0&919	&&	8&6e-4		&	3&3e-3	&	0&952	\\
	&		&	&	1			&	-3&2e-3	&	9&7e-3	&	0&924	&&	-9&3e-4	&	3&1e-3	&	0&953	\\
	&		&	&	10		&	-0&025	&	0&011	&	0&902	&&	-1&0e-3		&	4&5e-3	&	0&991	\\
\cline{3-17}
	&		&	5000	&	0	&	3&5e-4	&	9&9e-4	&	0&921	&&	-1&8e-4	&	3&1e-4	&	0&956	\\
	&		&	&	1				&	-2&7e-4	&	9&8e-4	&	0&922	&&	-1&4e-4	&	3&1e-4	&	0&950	\\
	&		&	&	10			&	-4&7e-3	&	1&0e-3	&	0&915	&&	-1&2e-4	&	4&5e-4	&	0&991	\\
\hline
Correctly	&	Mis-	&	50	&	0		&	0&025	&	0&10	&	0&934	&&	4&2e-3	&	0&030	&	0&954	\\
specified	&	specified	&			&	1		&	0&011	&	0&10	&	0&919	&&	1&8e-3	&	0&029	&	0&955	\\
	&		&			&	10	&	-0&18	&	0&28	&	0&933	&&	4&1e-3	&	0&030	&	0&952	\\
\cline{3-17}
	&		&	500	&	0	&	8&3e-3	&	9&9e-3	&	0&966	&&	-6&6e-4	&	3&0e-3	&	0&968	\\
	&		&	&	1			&	1&6e-3	&	9&8e-3	&	0&959	&&	8&0e-5		&	3&1e-3	&	0&965	\\
	&		&	&	10		&	-0&082	&	0&029	&	0&973	&&	-3&1e-4	&	2&9e-3	&	0&968	\\
\cline{3-17}
	&		&	5000	&	0	&	7&6e-3	&	1&1e-3	&	0&963	&&	2&9e-4		&	3&0e-4	&	0&967	\\
	&		&	&	1				&	1&8e-3	&	9&8e-4	&	0&964	&&	-1&8e-4	&	3&1e-4	&	0&967	\\
	&		&	&	10			&	-0&054	&	5&4e-3	&	0&916	&&	3&2e-4		&	2&9e-4	&	0&966	\\
\hline
\\
\end{tabular}
\end{center}
\end{table}

Results from this study are summarized in Table~\ref{t:one}. Coverage probabilities in correctly-specified-model settings with interference are quite comparable to those without interference. 
Coverage is generally better for the overall effect parameter than for the direct effect. This is likely a result of the probability weights being more stable for the overall effect parameter, due to the OERS being closer to the actual propensity score than the interventions assigning all subjects to exposure and no exposure. Double robustness merely implies consistency, so correct coverage is not necessarily expected in the settings with model mis-specification, though we do still see fairly decent coverage in these settings. In practice, we recommend estimating the nuisance functions under more adaptive models using machine learning tools such as super learner, so as to avoid model mis-specification, and preserve valid inference.

There is a general trend of bias increasing with $\beta$, likely due to the increase in variance of $Y$. MSE remains relatively stable across $\beta$ values, indicating that the conditional variance given $\bar{A}$ is stable with respect to $\beta$, and that bias is small relative to conditional variance. 
The double robustness property of our estimator is evidenced by the decrease in bias and MSE with sample size in each setting.


\section{Data Application}
\label{s:data}
Health care worker shortages are a widespread problem, particularly in sub-Saharan Africa, and pose serious public health challenges among populations with HIV and otherwise. 
To combat this issue, clinics have increasingly adopted task-shifting programs, in which patients identified as being at low risk for poor health outcomes have designated health care tasks shifted from a higher-cadre health care provider to a lower-cadre provider (e.g. from physician to mid-level provider; mid-level provider to nurse; nurse to lay worker; or in our case, mid-level provider to nurse) \citep{world2012taking}. 


Clearly, it is of great public health interest to understand the effects of task shifting, and in particular whether this practice is harmful for patients whose care is being shifted. If it is harmful, then this practice may not be advisable for widespread use; otherwise, it may be a viable option for mitigating health care worker shortages. Indeed, according to \cite{world2012taking}, ``Action on human resources for health is imperative if global commitments to the Millennium Development Goals, and to providing universal access to HIV services, are to be met. Action on task shifting is imperative as it provides the only realistic possibility of increasing the human resources fast enough to meet the urgent need.''

In their systematic review and meta-analysis of task-shifting studies, \cite{lassi2013quality} concluded, ``No difference between the effectiveness of care provided by mid-level health workers and that provided by higher-level health workers was found.'' However, they qualified these findings as being largely based on studies in developed countries, and highlighted a need for more studies conducted in Africa. 
In the LREC program, task-shifting was adopted in clinics in Kenya, and data were collected on eligible low-risk patients. As we described previously, an analysis of the data from this program was conducted in \cite{tran2016evaluating}, but the authors expressed concern that interference may have been present in this study due to the potential for the number of task-shifted patients to affect the time and quality of care nurses are able to provide. We apply the methodology developed in this paper, which accounts for this form of interference, to study the effects of task shifting. 

We conduct our analysis on these data independently within each clinic, estimating various effects using the TMLE defined in Section \ref{s:estimation}. 
The nuisance functions $g^*$ and $\bar{Q}$ are estimated using a super learner with a library consisting of the marginal mean, generalized linear models, LASSO, random forest, generalized boosted models, and support vector machines.

\begin{figure}
\centering
\includegraphics[width=.9865\textwidth]{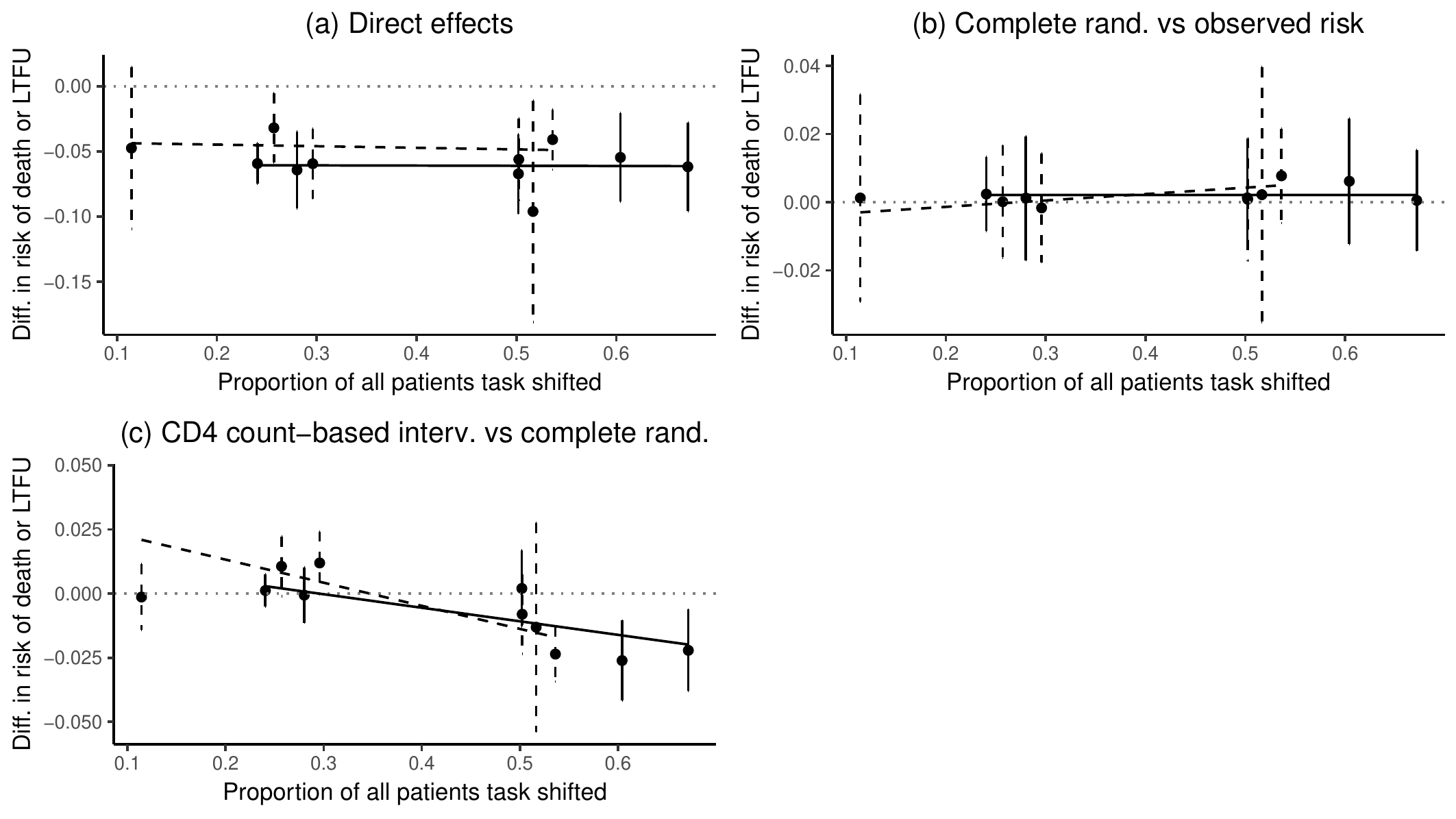}
\caption{(a) The estimated direct effects of task shifting on risk of death or loss to follow up in each clinic when the observed proportion of all patients task shifted is held fixed. (b) The estimated overall effect of a complete randomization intervention on risk of death or loss to follow up compared to the observed proportion of cases. (c) The estimated overall effect comparing an intervention task shifting only the $S_n$ patients with the highest CD4 counts to complete randomization on risk of death or loss to follow up. 
The vertical bars represent the corresponding 95\% confidence intervals. Solid lines indicate results for urban clinics; dashed lines indicate results for non-urban clinics. Estimates are plotted against the proportion of all patients task-shifted in each clinic. The lines spanning the plot horizontally represent the fitted MSMs, summarizing how these effects vary as functions of the proportion of patients task-shifted given the urban designation of the clinic.}
\label{f:results}
\end{figure}

Our results are summarized in Figure \ref{f:results}. In addition to the individual clinic-level effect estimates, each panel depicts an estimated marginal structural model (MSM) summarizing each effect across the different clinics as a function of the proportion of all patients task shifted. We relegate discussion of the MSMs to Web Appendix A. 
The plot in panel (a) depicts the estimated direct effects of task shifting in settings in which only the observed number of patients are task shifted. 
We observe significant effects in all but one clinic, suggesting that task-shifting is beneficial in these clinics. The only non-significant effect estimate was found in the clinic with the lowest proportion of patients task-shifted.


It is also of interest to compare the effects of various ERSs. The plots in panels (b) and (c) depict estimates of such contrasts. In panel (b) complete randomization is contrasted with the observed risk of death or loss to follow up. This effect contrast essentially tells us how different outcomes would have been if the observed number of task-shift assignments had been allocated randomly (in contrast to the observed non-random assignment mechanism). No effect is statistically significant among any of the clinics. 
The contrast depicted in panel (c) is between the ERS assigning the $S_n$ patients with the highest CD4 counts to be task shifted compared with complete randomization. If one were to believe that task shifting is most beneficial to healthy patients, then using CD4 count as a proxy for patients' health status, one might be interested in this intervention, which corresponds to only task shifting the healthiest $S_n$ patients. The effect contrast essentially tells us how much outcomes could be improved if task shifting were assigned according to CD4 count over an intervention completely ignoring covariate information. Three of these clinics have statistically-significant effect estimates. This suggests an improvement in outcomes under the CD4 count-based intervention in these particular clinics.

\section{Discussion}
\label{s:discussion}
In this paper, we have developed new methodology to perform causal inference in settings with stratified interference when the number of causally-separated groups is small or only one, but the sample sizes of these groups are moderate to large. A central challenge in the setting of a single causally-dependent group under stratified interference is that the statistical support for the outcome regression is limited to a single observed proportion of subjects receiving the exposure. Our proposed methodology overcomes this by focusing on identification and estimation of causal parameters that summarize the distribution of counterfactual outcomes under interventions that preserve the observed proportion of subjects exposed.

Our proposed methodology does not require the implementation of a new estimator, but rather utilizes the existing TMLE developed for the average counterfactual outcome 
in the absence of interference. In the case of the overall effect of an exposure reallocation scheme, one need only derive the appropriate marginal intervention rule from the user-specified joint rule; otherwise, no additional work is required. Crucially, our methodology is doubly robust, which permits the use of data-adaptive/machine learning tools to estimate the response surface and propensity score.

Our methodology is particularly relevant to situations in which the efficacy of an exposure varies with the proportion of subjects receiving it. This may often occur when the exposure is a limited, shared resource, such as in our task-shifting data example, where health care workers' time is the shared resource. Task shifting is a potentially widely useful approach to combat health care worker shortages worldwide, thus our methodology may be quite useful in other such studies.

In our data application, we observed a statistically-significant improvement in risk of death or loss to follow up as a result of shifting patients from clinical officers to nurses, provided the proportion of patients task shifted remains fixed, in all but one clinic. This finding is consistent with the results reported by \cite{tran2016evaluating} in their analysis of longitudinal interventions on enrollment, in which they accounted for time-dependent confounding, but not potential interference. 
While this result may reflect a truly beneficial effect, due for example to improved quantity or quality of care that nurses are able to provide, it may also reflect unmeasured confounding, with patients at lower risk in ways not captured by baseline covariates more likely to be enrolled in the task-shifting program \citep{tran2016evaluating}.

We have shown how to estimate the overall effects of interventions within the class of exposure reallocation schemes, which reallocate the observed number of exposure assignments based on some user-specified rule. If the effect of the individual-level exposure varies depending on an individual's covariates, such reallocation schemes have the potential to improve outcomes by assigning the same number of interventions to the individuals most likely to benefit. For instance, it may be that healthier patients respond better to task shifting, in which case an exposure reallocation scheme based on CD4 count might be preferable to complete randomization, as we have demonstrated evidence of in some clinics. (In practice, the eligibility criteria include CD4 count, and the observed task-shifting decisions used some subsequent assessment of illness beyond that initial cut off.) Even in this case, however, task-shifting might still be beneficial for sicker patients as well, in which case it is possible that we should be task-shifting more than just the observed number of patients. Yet, without additional causal assumptions and observations from other groups with higher task-shifting rates, the effect of task shifting more patients cannot be estimated without extrapolating beyond the support of the observed data. Therefore, assertions regarding effects under interventions that task shift more than the observed proportion of patients must be interpreted with care. A natural follow-up question is: Which intervention in the class of exposure reallocation schemes is best? We plan to pursue estimation of this optimal intervention in future work.

The proposed methodology has several limitations. The inability to estimate causal parameters under interventions in which a different proportion of subjects is exposed than is actually observed 
is a limitation inherited directly from the data. When subjects are enlisted over an extended period of time, our methodology can only be applied to a subset of subjects enlisted within a limited time frame, as was done in our data example, rather than the entire population enlisted over the duration of the study. While we allow for subjects' exposures to affect one another's outcomes, we do not account for subjects' covariates affecting one another's exposures or outcomes. 
Finally, our methodology relies on the stratified interference assumption, the plausibility of which will vary widely from application to application. Simply because one lacks knowledge of the underlying structure of interference, does not mean one should blithely assume stratified interference. However, as we have argued, we believe it can be a reasonable approximation, especially in settings in which interference is induced by an exposure being a limited resource being shared among subjects.

\section*{Acknowledgments}

Research reported in this publication was supported by the National Institute Of Allergy And Infectious Diseases (NIAID), Eunice Kennedy Shriver National Institute Of Child Health \& Human Development (NICHD), National Institute On Drug Abuse (NIDA), National Cancer Institute (NCI), and the National Institute of Mental Health (NIMH), in accordance with the regulatory requirements of the National Institutes of Health under Award Number U01AI069911East Africa IeDEA Consortium. The content is solely the responsibility of the authors and does not necessarily represent the official views of the National Institutes of Health. This research has been supported by the President's Emergency Plan for AIDS Relief (PEPFAR) through USAID under the terms of Cooperative Agreement No. AID-623-A-12-0001. It is made possible through joint support of the United States Agency for International Development (USAID). The contents of this article are the sole responsibility of AMPATH and do not necessarily reflect the views of USAID or the United States Government. We would like to acknowledge the contributions of Drs. Abraham Siika, Sylvester Kimaiyo, AMPATH Eldoret Kenya and Dr. Kara Wools-Kaloustian, Indiana University School of Medicine, in designing the LREC program and Dr. Constantin Yiannoutsos and Ms. Beverly Musick from the East Africa IeDEA Regional Data Center, Indiana University as well as the AMPATH patients that contributed data for this project.\vspace*{-8pt}

\appendix

\newpage

\section{Effect modification and joint effects with $\bar{A}$ when observing multiple groups under partial interference}
\label{s:spillover}
Suppose one has data from multiple groups, between which there is assumed to be no interference, and the stratified interference assumption is believed to hold within each group. In this case, there will be statistical support for causal parameters depending on a range of $k_n(\bar{A})$, hence we will be able to study effect modification by $k_n(\bar{A})$.

We propose a method for doing so that involves first estimating any of the parameters defined in Section 2 within each group independently, and then positing a marginal structural model for these effects as a function of $k_n(\bar{A})$ and group-level data, such as summary statistics of covariates within each group. 
In the LREC example, we can study how the individual-level intervention is modified by the proportion of all patients exposed by estimating how the direct effect of task shifting varies as the proportion of patients task-shifted takes different levels. Asymptotics for this method rely on the sample sizes within each group, rather than the number of groups. 

MSMs are working models for counterfactual quantities given a summary measure of the exposure and possibly effect modifiers. While they were originally defined for the purpose of estimating causal effects due to repeated exposures in the presence of time-varying confounding \citep{robins1998marginal}, their usage has since been expanded. Here we define MSMs for the mean of group-level effects as a function of the summary function of the observed proportion exposed as well as (potentially) group-level covariate data. Without any additional assumptions, we consider the proportion exposed to be a potential effect modifier, rather than an intervention variable.

For groups $j=1,\ldots,J$, define $\bar{Y}^{*}_j\equiv n_j^{-1}\sum_{i=1}^{n_j}Y_{ij}^{*}$, where $Y_{ij}^{*}$ can be either of the counterfactual outcomes defined in Section 2 of subject $i$ in group $j$ and $n_j$ is the number of subjects in group $j$, and let $\psi_{0,j}\equiv E(\bar{Y}_j^{*}\mid W^{n_j})$. Posit a parametric working MSM $E\{\psi_{0,j}\mid V_j\}=m(V_j;\beta)$ in terms of a finite-dimensional parameter $\beta$, where $V_j\allowbreak\equiv\allowbreak \{k_{n_j}(\bar{A}_j),\allowbreak d_{n_j}(W^{n_j}),\allowbreak G_j\}$, $d_{n_j}$ is some user-specified, potentially vector-valued summary function of covariates of interest within group $j$, and $G_j$ contains group-level data common to all subjects in group $j$. The latter two elements of $V_j$ are optional, and should be selected parsimoniously if the number of groups is relatively small, so as not to overfit the model. In general, the MSM is a working model, and is not necessarily to be believed, but rather provides a model for summarizing the relationship between $\psi_{0,j}$ and $V_j$.

In the LREC example, we let $G_j$ be an indicator of whether clinic $j$ is located in an urban area, and let $d_{n_j}$ be empty. For each of the effects we study, we project the clinic-level effects, say $\psi_j$, onto the marginal structural model $m(V_j;\beta)=\beta_0+\beta_1 G_j+\beta_2 k_{n_j}(\bar{A}_j)+\beta_3 G_j\times k_{n_j}(\bar{A}_j)$, and we choose $k_{n_j}$ to be the linear function mapping $\bar{A}_j$ to the overall proportion of all patients in clinic $j$ task-shifted at baseline. 


The interpretation of the MSM thus far has been somewhat nontraditional; typically an MSM summarizes how a counterfactual mean varies as a function of an intervention, whereas we have considered the intervention to be fixed, and the MSM has summarized how the counterfactual mean for a single intervention varies as a function of $k_n(\bar{A})$. We now consider conditions under which the MSM can be interpreted as a model that is causal with respect to $k_n(\bar{A})$. Let $\psi_j(k)$ be the counterfactual value $\psi_j$ would take if, possibly contrary to fact, $\bar{A}_j$ were set to a level $\bar{a}_j$ such that $k_{n_j}(\bar{a}_j)=k$. When $\psi_j(k)\ci\bar{A}_j\mid \{d_{n_j}(W^{n_j}),G_j\}$ for all $k$, i.e., $\{d_{n_j}(W^{n_j}),G_j\}$ controls for confounding between $\bar{A}_j$ and $\psi_j(k)$, then the MSM can further be interpreted as a joint causal model under the individual-level intervention used to define $\psi_j$ and an intervention on $k_{n_j}(\bar{A}_j)$. When $\psi_j$ is the average counterfactual outcome under the intervention assigning all subjects to no exposure, this MSM will describe the spillover effect of the intervention on non-exposed subjects. Of course, when the number of causally-disconnected groups is relatively small, as in our data example, there is unlikely to be adequate balance in confounders at various levels of $k_{n_j}(\bar{A}_j)$, and hence this causal interpretation will not be very reliable. Like other methods proposed for estimating spillover effects \citep{hudgens2008toward,tchetgen2012causal}, ours will be most appropriate when the number of groups is moderate to large.

We now discuss estimation of the MSM, which is the same regardless of interpretation. Define a loss function, $L\{m(V_j;\beta),\psi_j\}$, for $\beta$, for instance the squared-error loss or logistic loss. Then the true $\beta_0$ is defined to be $\argmin\limits_{\beta}\sum_{j=1}^Jw_jL\{m(V_j;\beta),\psi_{0,j}\}$, where $w_j$ is some user-specified weight, such as the inverse variance of the estimated $\hat{\psi}_j$, which we use in the LREC example, or potentially simply one for all $w_j$. Thus, $\beta_0$ is well defined as a projection of $\psi_{0,j}$ onto the working model, regardless of whether it is correctly specified. The MSM parameter is estimated by plugging in the estimated $\hat{\psi}_j$, and minimizing with respect to $\beta$, i.e., $\hat{\beta} = \argmin\limits_{\beta}\sum_{j=1}^J w_jL\{m(V_j;\beta),\psi_{n_j,j}\}$. Define $S(\beta,\psi^J)\equiv\sum_{j=1}^Jw_j\nabla_{\beta}L\{m(V_j;\beta),\psi_j\}$, where $\psi^J$ is the vector of $\psi_j$ for all $j$. Then $S(\beta_0,\psi_0^J)=S(\hat{\beta},\hat{\psi}^J)=0$, and applying Taylor expansions to $S(\hat{\beta},\hat{\psi}^J)=0$ about $\beta_0$, then $\psi_0^J$, successively, yields 
\[\sqrt{n}(\beta_n -\beta_0)\approx -\left\{\nabla_{\beta}S(\beta,\psi_0^J)\rvert_{\beta_0}\right\}^{-1}\nabla_{\psi^J}S(\beta_0,\psi^J)\rvert_{\psi^J_0}\left[\begin{array}{c}
\sqrt{n/n_1}\sqrt{n_1}\left\{\hat{\psi}_1 -\psi_{0,1}\right\}\\
\vdots\\
\sqrt{n/n_J}\sqrt{n_J}\left\{\hat{\psi}_J -\psi_{0,J}\right\}
\end{array}\right],\]
and asymptotic normality follows from asymptotic normality of the estimates $\hat{\psi}^J$, provided $\lim_{n\rightarrow\infty}\allowbreak n\allowbreak /\allowbreak n_j$ exists for each $j$. Thus, $\hat{\beta}$ is a linear combination of the independent estimators $\hat{\psi}_j$, and Wald tests and confidence intervals for $\hat{\beta}$ can be constructed using the point estimates, $\hat{\psi}_j$, and their variance estimates. When the MSM is linear in $\beta$ and inverse variance weights are used, $\hat{\beta}=(XWX^T)^{-1}XW\hat{\psi}^J$ and $(XWX^T)^{-1}$ consistently estimates its variance, where $X$ is the design matrix for the MSM, and $W$ is a diagonal matrix with the inverse variance of $\hat{\psi}_j$ on the $j$th diagonal for each $j$.


For each of the three effects studied in our data analysis, the MSM slope estimates for $k_n(\bar{A})$ differed very little between urban and non-urban clinics. Additionally, none of these slope estimates were statistically significant, meaning we lack evidence to conclude that the overall proportion of patients task-shifted modifies any of these effects. However, for the sake of demonstration, we provide an interpretation of the slope of the overall effects comparing an intervention based on CD4 count to complete randomization with respect to overall task-shift coverage among urban clinics. Were this negative slope estimate statistically significant, we would conclude that the the CD4 count-based intervention becomes more favorable compared with complete randomization as the overall task-shift coverage within the clinic increases.

\section{Approximate exposure reallocation schemes}
There exists a larger class of data-adaptive interventions under which inference for overall effect parameters is possible, which we term \emph{approximate exposure reallocation schemes} (AERS). Specifically, this class contains all interventions with individual-level marginal probabilities $g^*_{\theta_n}$ such that $n^{-1}\sum_{i=1}^n\sum_{a=0}^1ag^*_{\theta_n}(a\mid W_i)=\bar{A}$. These need not preserve the exact proportion of subjects exposed, but rather have individual-level conditional probabilities of exposure given $W^n$ equal to the observed proportion exposed, and generate $A^{*n}$ with sample mean that will converge to the observed $\bar{A}$ asymptotically. One example of an AERS is the Bernoulli intervention that assigns each subject to be exposed with probability $\bar{A}$.  An ERS is guaranteed to produce marginal individual-level probabilities $g^*_{\theta_n}$ satisfying the AERS criterion, hence the class of AERSs contains the class of ERSs.

The following result shows that the causal nonparametric identification formula for $\psi^{g^*}$ under an AERS is asymptotically equivalent to $\Psi^{g^*\!,\bar{A}}(\bar{Q}_0)$ on the order of $n^{-1/2}$, such that asymptotic inference for $\Psi^{g^*\!,\bar{A}}(\bar{Q}_0)$ is valid for $\psi^{g^*}$.
\begin{theorem}
Suppose (i) the NPSEM defined in (1) contains the true set of underlying counterfactual distributions, (ii) $\bar{Q}_0$ is differentiable in $\bar{A}$ such that the limit of its derivative is a bounded function of $\bar{A}$, (iii) Assumption 1 holds for $\pi = \bar{A}$, and (iv) Assumption 2 holds. Then if $g^*$ is an AERS,
\[\psi^{g^*}=\frac{1}{n}\sum_{i=1}^n\sum_{a^n}\bar{Q}_{0}\{W_i,a_i,k_n(\bar{a})\}g^*(a^n\mid W^n) = \Psi^{g^*\!,\bar{A}}(\bar{Q}_0) + o_p(n^{-1/2}).\]
\end{theorem}
Thus, when $g^*$ is an AERS, 
the causal identifying functional for the overall effect parameter $\psi^{g^*}$ in Theorem 1 is asymptotically equivalent to $\Psi^{g^*\!,\bar{A}}(\bar{Q}_0)$ on the order of $n^{-1/2}$, and inference for $\psi^{g^*}$ is asymptotically equivalent to that for $\Psi^{g^*\!,\bar{A}}(\bar{Q}_0)$. 

\section{Review of TMLE under no interference}
The analogous population-average causal target parameter under $\mathcal{M}_{iid}$ is $\psi_{iid}^p \equiv E(Y^{g^*})$, where $Y^{g^*}\equiv f_Y(W,A^*,U^Y)$ and $A^*$ is the exposure assigned under intervention $g^*$. Letting $Q\equiv\{\bar{Q},Q_W\}$, the parameter $\psi_{iid}^p$ is identified by $\Psi_{g^*}(Q_0)\equiv\int_w\sum_{a=0}^1\bar{Q}_0(a,W)g^*(a\mid W)dQ_{W,0}(w)$. The PATE is defined as the contrast in $\psi_{iid}^p$ comparing the interventions $g^*(1\mid W)=1$ and $g^*(0\mid W)=1$.

A semiparametric efficient estimator for $\psi_{iid}^p$ can be obtained by solving the efficient influence curve estimating equation. For $\Psi^{g^*}(Q)$ under $\mathcal{M}_{iid}$ the efficient influence curve is
\[D_{g^*}^p(Q_0,g_0)(O_i)\equiv \frac{g^*(A_i\mid W_i)}{g_0(A_i\mid W_i)}\left\{Y-\bar{Q}_0(A_i,W_i)\right\}+\sum_{a=0}^1\bar{Q}_0(a,W_i)g^*(a\mid W_i)-\Psi_{g^*}(Q_0),\]
and the sample analog to the identity $P_{iid,0}D_{g^*}^p(Q_0,g_0)=0$ forms the efficient influence curve estimating equation: $P_nD_{g^*}^p(Q,g)=0$. An estimating equation approach solves this equation for $\Psi_{g^*}(Q)$ in the parameter space, yielding the estimator
\[\frac{1}{n}\sum_{i=1}^n\left[\frac{g^*(A_i\mid W_i)}{g_n(A_i\mid W_i)}\left\{Y-\bar{Q}_n(A_i,W_i)\right\}+\sum_{a=0}^1\bar{Q}_n(a,W_i)g^*(a\mid W_i)\right],\]
where $\bar{Q}_n$ and $g_n$ are estimates of the nuisance parameters $\bar{Q}_0$ and $g_0$, respectively. This particular estimator is known as an augmented inverse probability weighting (A-IPW) estimator. By contrast, construction of the TMLE involves solving the efficient influence curve equation for $(Q,g)$ in the model space. The TMLE is the estimator that results from plugging in the $Q$ portion of this solution into $\Psi_{g^*}$, hence it is a substitution estimator. Thus, the TMLE has the properties of being both semiparametric efficient, since it solves the efficient influence curve estimating equation, and having added stability and respecting the boundaries of the parameter space, since it is a substitution estimator.

We now describe the algorithm by which the solution $(Q^*_n,g_n)$ to $P_nD_{g^*}^p(Q,g)=0$ is obtained. The algorithm takes as input an initial consistent estimate of $(Q,g)$, say $(Q^0_n,g^0_n)$, and outputs $Q^*_n$. Machine-learning-based initial estimates that are optimal for $Q$ will generally produce an overly-biased estimate when plugged into $\Psi_{g^*}$, hence a so-called targeting step for $\bar{Q}^0_n$ is required. For our particular problem, updating $g^0_n$ is in fact unnecessary; the final plug-in estimator does not depend on $g$, and there is no iteration required. Updating $Q_W^0$ is generally never required when the empirical estimator is used for $Q^0_W$. The empirical estimator is the nonparametric maximum likelihood estimator, and will hence not generate bias when plugged into $\Psi_{g^*}$.

The targeting step for $\bar{Q}$ involves formulation of a parametric submodel and a loss function that are compatible in a particular manner. We will consider a loss function, $L(\cdot)$, that only depends on $P_{iid}$ via $\bar{Q}$, and satisfies $\bar{Q}_0=\argmin_{\bar{Q}}P_{iid,0}L(\bar{Q})$. We can then express our submodel as $\{\bar{Q}(\epsilon):\epsilon\}$ where $\epsilon$ is a scalar. The objective is to define a submodel--loss function pair such that when the submodel is centered at the solution to $P_nD_{g^*}^p(Q,g)=0$, i.e., $\bar{Q}(0)=\bar{Q}^*$, the solution will also minimize the loss function.

By iteratively minimizing the loss function on a submodel centered on the latest update, one will converge to the solution, which we denote $\bar{Q}^*$. For a loss function that is differentiable in $\epsilon$, this occurs if we define $L$ and $\{\bar{Q}(\epsilon):\epsilon\}$ such that $\frac{d}{d\epsilon}L\{\bar{Q}(\epsilon)\}\vert_{\epsilon=0}=D^p_{g^*}(Q,g)$. To this end, using a binary $Y$ as an example, we will employ the log-likelihood loss function (hence the ``maximum likelihood'' in TMLE), $L(Q)$: $L(\bar{Q})(O)=-\log\bar{Q}(W,A)^Y\{1-\bar{Q}(W,A)\}^{1-Y}$. Further, we will use the logistic submodel: $\logit\bar{Q}^0_n(\epsilon)(W,A)=\logit\bar{Q}^0_n(W,A)+\epsilon H_n^*(W,A)$, where $H_n^*(W,A)\equiv g^*(A\mid W)/g_n(A\mid W)$, which is often referred to as a ``clever covariate'' due to it being the correct choice of covariate in this submodel that yields the desired property $\frac{d}{d\epsilon}L\{\bar{Q}(\epsilon)\}\vert_{\epsilon=0}=D^p_{g^*}(Q,g)$. This loss function-submodel pairing is also recommended for a continuous $Y$ rescaled to the $[0,1]$ interval \citep{gruber2010targeted}.

The algorithm then proceeds by finding the optimal $\bar{Q}(\epsilon)$ in the submodel with respect to $L$. This is done by finding $\epsilon_n=\argmin_{\epsilon}\sum_{i=1}^nL\{Q^0_n(\epsilon)(O_i)\}$, which can be accomplished by simply fitting a logistic regression of $Y$ on the clever covariate, $H_n^*(W,A)$, with the intercept replaced by an offset, namely $\logit\bar{Q}^0_n(W,A)$. The updated $\bar{Q}$ estimate is then $\bar{Q}^*_n(W,A):=\expit\{\logit\bar{Q}^0_n(W,A)+\epsilon_nH^*_n(W,A)\}$. For this particular parameter, the algorithm converges after the first update, and the updated $\bar{Q}^*_n$ is the estimate that is plugged into $\Psi_{g^*}$ for the final estimate of $\psi^p_{iid}$. 

Suppose the following regularity conditions hold (i) $\frac{g^*(A\mid W)}{g_0(A\mid W)}<\infty$ a.e.,
(ii) $D^p_{g^*}(\bar{Q}^*_n, g_n)$ belongs to a Donsker class with probability approaching one, and (iii) \[P_0\left\{\bar{Q}_n^*(W,A)-\bar{Q}_0(W,A)\right\}^2P_0\left\{g_n(A\mid W)-g_0(A\mid W)\right\}^2=o(n^{-1/2}).\]
Then the resulting TMLE is asymptotically normal with asymptotic variance equal to $P_{iid,0}\allowbreak D^p_{g^*}(\allowbreak Q_0,\allowbreak g_0)^2$, which can be very readily estimated by $\frac{1}{n}\sum_{i=1}^nD^p_{g^*}(\bar{Q}^*_n,g_n)(O_i)^2$. The Donsker condition can be relaxed if one instead uses cross validation in the targeting step of the above algorithm, which is known as CV-TMLE \citep{zheng2011cross}.

\cite{balzer2016targeted} show that the TMLE for the PATE contrasting the static, deterministic interventions $g^*(A_i=a^*\mid W_i)=I(A_i=a^*)$ for $a^*\in\{0,1\}$ is also a consistent and asymptotically normal estimator for the conditional average treatment effect (CATE) contrasting these interventions, i.e., $\psi^c_{iid}\equiv \Psi^{g^*}(\bar{Q}_0,Q_{W,n}) = \frac{1}{n}\sum_{i=1}^n\sum_{a=0}^1\bar{Q}_0(a,W_i)g^*(a\mid W_i)$. The only difference is that its influence curve is instead \[D^c_{g^*}(\bar{Q}_0,g_0)(O_i)\equiv\frac{g^*(A_i\mid W_i)}{g_0(A_i\mid W_i)}\left\{Y-\bar{Q}_0(A_i,W_i)\right\}.\] They show that $D^c_{g^*}$ is uncorrelated with the dropped portion of $D^p_{g^*}$, i.e., $\sum_{a=0}^1\bar{Q}_0(a,W_i)g^*(a\mid W_i)-\Psi_{g^*}(Q_0)$, and hence the same TMLE has reduced asymptotic variance when used to estimate $\psi^c_{iid}$. Specifically, the asymptotic variance reduces to $P_{iid,0}\{D^c_{g^*}(\bar{Q}_0,g_0)^2\}$, which can also be estimated by its sample analog with plugged-in estimated nuisance functions. This result extends analogously to dynamic and stochastic interventions.

\section{Population-average effects}
\label{s:population}
In the main article, we only considered conditional-average effects given the observed covariates. However, population-average effects may also be of interest, particularly if one wishes to generalize results to another group of subjects whose covariates are not yet observed, and who are thought to be sampled from the same superpopulation as the observed subjects. 
Our population-average parameter of interest is now $\psi_p^{g^*}\equiv E(\bar{Y}^{g^*\!,\bar{A}})$. ERSs will typically result in individual-level interventions that depend on the covariates of other subjects. Since we now wish to estimate parameters that marginalize over the distribution of all subjects' covariates, asymptotic normality of the estimator defined in Appendix C may only hold for a further restricted class of interventions, which are somewhat unnatural choices of interventions. Therefore, in this section we shift our focus to individual-level interventions $g^*_{\theta_n}(A_i\mid W_i)$ that do not depend on the covariates of any other subject.

The parameter $\psi_p^{g^*}$ under such an intervention will not correspond to an overall effect, since no restrictions can be enforced on the individual-level intervention $g^*$ to ensure the vector $A^{*n}$ has sample mean approximately equal to the observed $\bar{A}$ without the remaining subjects' covariate values. The parameters contrasted in the direct effect are examples of $\psi_p^{g^*}$ under an individual-level intervention. 
The following result states that for this parameter, the estimator defined in Appendix C remains consistent and asymptotically normal, but with a larger asymptotic variance.
\begin{theorem}
Suppose the conditions in Appendix F all hold, $k_n(\bar{A})$ has limit $k_0$ as $n\rightarrow\infty$, $g^*$ is an individual-level intervention common to each subject $i$, and depending only on $W_i$ for each subject $i$. Then the TMLE $\psi_n=\Psi^{g^*}(\bar{Q}^*_n,Q_{W,n})$ for $\psi_p^{g^*}$ under $\mathcal{M}_{iid}$ is a consistent and asymptotically normal estimator for $\psi_p^{g^*}$ under $\mathcal{M}_{si}$, with asymptotic variance $\sigma^2=\sigma_Y^2+\sigma_W^2$, where \[\sigma_W^2\equiv E_{p_0}\left\{\sum_{a=0}^1\bar{Q}_0(W,a,k_0)g^*(a\mid W)-\psi_p^{g^*}\right\}^2.\]
\end{theorem}
The condition that $k_n(\bar{A})$ converges precludes $k_n(\bar{A})=n\bar{A}$, whereas this was permitted under Theorem 3.

Estimation and inference for population-average versions of parameters corresponding to overall effect parameters is considerably more challenging due to the dependence of the individual-level intervention $g^*_{\theta_n}$ on the covariates of other subjects inherent to ERSs (apart from complete randomization). Therefore, we leave this problem open for future work.

\section{Simulation study details}
\label{s:sims}
Our simulation design combines different sample sizes (50, 500, and 5000) and interference levels governed by the parameter $\beta$ (0,1,
10). In each setting, we draw 5000 samples from the following data generating mechanism: $W_i \sim N(0,1)$, $\p (A_i=1\mid W_i) = \expit W_i$, $Y_i^{0,\bar{A}} \sim N(0,1)$, $Y_i^{1,\bar{A}} = Y_i^{0,\bar{A}}+W^n(1-\beta\bar{A})$, $Y_i = A_iY_i^{1,\bar{A}}+(1-A_i)Y_i^{0,\bar{A}}\;\; \forall i\in\{1,\ldots,n\}$. The structural equation for $Y$ is then $f_Y(W_i,A_i,\bar{A},U^Y_i)=A_iW_i(1-\beta\bar{A})+U^Y_i$, where $U^Y_i=Y_i^{0,\bar{A}}\sim N(0,1)$, and so the form of interference is linear in $\bar{A}$, 
and its slope coefficient $\beta$ determines the strength of the interference. The setting with $\beta=0$ corresponds to there being no interference. The setting with $\beta=1$ corresponds to a setting in which shifting everyone would completely negate the effect of the exposure. When $\beta>1$, the direction of the individual exposure effect reverses when enough individuals are exposed. For example, in the LREC program as the number of patients enrolled in the task-shifting program increases, the workload of clinical officers is reduced, potentially improving the care of patients both enrolled and unenrolled in the program. At the same time, the more patients enrolled, the greater the demand is on nurses and other care providers. Thus, beyond some threshold enrollment, increasing the number of patients enrolled may result in worse outcomes for enrolled patients. The settings in which $\beta$ is 
10 thus allows us to examine how well our method works under particularly strong interference.

We consider two data-adaptive causal parameters depending on the observed $\bar{A}$: (i) the direct effect when $\bar{A}$ of subjects are exposed, and (ii) the overall effect under the ERS that assigns subjects with the $S_n\equiv\sum_{i=1}^nA_i$ highest $W$ values to exposure: $g^*(a^n\mid w^n)=\prod_{i=1}^nI(w_i>w_{(n-S_n)})^{a_i}I(w_i\leq w_{(n-S_n)})^{1-a_i}$, where $w_{(i)}$ is the $i$th largest value among $w^n$. The empirical means and standard deviations of the data-adaptive causal parameters under each simulation setting are displayed in Table~\ref{t:target_params}.
\begin{table}
\begin{center}
\caption{Target parameter summary statistics}
\label{t:target_params}
\centering
\begin{tabular}{l c r@{.}l r@{.}l c r@{.}l r@{.}l}
\\
\hline\hline
&&	\multicolumn{9}{c}{Target parameter}	\\
\cline{3-11}
&&	\multicolumn{4}{c}{Direct effect}	&&	\multicolumn{4}{c}{OERS}\\
\cline{3-6} \cline{8-11}
$n$	&	$\beta$ &\multicolumn{2}{c}{Mean}&\multicolumn{2}{c}{SD}&&\multicolumn{2}{c}{Mean}&\multicolumn{2}{c}{SD}\\
\hline
50		&	0		&	-1&1e-3	&	0&14		&&	0&39	&	0&083\\
			&	1		&	-3&0e-3	&	0&071		&&	0&19	&	0&040\\
			&	10	&	-0&032	&	0&58		&&	-1&6	&	0&50\\
500		&	0		&	1&1e-3	&	0&045		&&	0&40	&	0&026\\
			&	1		&	-2&4e-4	&	0&023		&&	0&20	&	0&013\\
			&	10	&	-4&9e-3	&	0&18		&&	-1&6	&	0&16\\
5000	&	0		&	-8&8e-5	&	0&014		&&	0&40	&	8&3e-3\\
			&	1		&	-6&9e-5	&	7&2e-3	&&	0&20	&	4&1e-3\\
			&	10	&	-6&7e-4	&	0&057		&&	-1&6	&	0&051\\
\hline
\\
\end{tabular}
\end{center}
\end{table}

For this data generating process, the intervention assigning the exposure to subjects with the highest values of $W$ corresponds with the optimal exposure reallocation scheme (OERS) within the class of all ERSs. When $\beta<\bar{A}^{-1}$, this is the OERS when a larger $Y$ is considered more favorable; otherwise it is the OERS when a smaller $Y$ is more favorable. In our setting, the former will always be the case for $\beta=1$, while the latter will almost always be the case when $\beta$ is 
10. We regard this rule as known, rather than estimated. In future work, we hope to develop theoretical results for estimating the OERS. As for now, it is of interest to show that effects under such an optimal rule, once known, can be estimated well.

We consider three different parametric models for estimation: one in which the outcome regression is correctly specified, but the propensity score is not, one in which the reverse is true, and one in which both are correctly specified. The correctly-specified model for the outcome regression is $\bar{Q}(W,A;\beta)=\beta_0+\beta_1W+\beta_2A+\beta_3WA$; the mis-specified model has no interaction term: $\bar{Q}(W,A;\beta^*)=\beta_0^*+\beta_1^*W+\beta_2^*A$. The correctly-specified model for the propensity score is $g(1\mid W;\gamma)=\expit\left(\gamma_0+\gamma_1W\right)$; the mis-specified model uses a probit link: $g(1\mid W;\gamma^*)=\Phi\left(\gamma_0^*+\gamma_1^*W\right)$, where $\Phi$ is the inverse probit function.

\begin{table}
\begin{center}
\caption{Simulation results for the direct effect and overall effect under the optimal exposure reallocation scheme (OERS) based on 5000 samples. MSE is mean squared error, and CP is the Monte Carlo coverage probability of the 95\% confidence interval.}
\label{t:one}
\centering
\begin{tabular}{c c l c r@{.}l r@{.}l r@{.}l c r@{.}l r@{.}l r@{.}l}
\\
\hline\hline
&&&&	\multicolumn{13}{c}{Target parameter}	\\
\cline{5-17}
&&&&	\multicolumn{6}{c}{Direct effect}	&&	\multicolumn{6}{c}{Overall effect -- OERS}\\
\cline{5-10} \cline{12-17}
$\bar{Q}$	&	$g$	&	$n$	&	$\beta$ &\multicolumn{2}{c}{Bias}&\multicolumn{2}{c}{MSE}&\multicolumn{2}{c}{CP}&&\multicolumn{2}{c}{Bias}&\multicolumn{2}{c}{MSE}&\multicolumn{2}{c}{CP}\\
\hline
Correctly	&	Correctly	&	50	&	0		&	0&010	&	0&10	&	0&916	&&	2&2e-4	&	0&027	&	0&948	\\
specified	&	specified	&	&	1		&	4&6e-3	&	0&10	&	0&917	&&	8&4e-5	&	0&027	&	0&951	\\
	&		&	&	10	&	-0&17	&	0&16	&	0&837	&&	0&026		&	0&028	&	0&942	\\
\cline{3-17}
	&		&	500	&	0		&	2&0e-3	&	9&7e-3	&	0&955	&&	3&0e-4	&	2&7e-3	&	0&960	\\
	&		&	&	1		&	-1&8e-3	&	9&6e-3	&	0&958	&&	1&4e-3	&	2&7e-3	&	0&962	\\
	&		&	&	10	&	-0&026	&	0&011	&	0&942	&&	5&7e-3	&	2&7e-3	&	0&960	\\
\cline{3-17}
	&		&	5000	&	0		&	1&1e-4	&	9&7e-4	&	0&958	&&	-1&5e-4	&	2&6e-4	&	0&964	\\
	&		&	&	1		&	8&6e-5	&	9&6e-4	&	0&961	&&	6&9e-5		&	2&6e-4	&	0&965	\\
	&		&	&	10	&	-3&0e-3	&	9&8e-4	&	0&961	&&	6&3e-4		&	2&7e-4	&	0&960	\\
\hline
Mis-	&	Correctly	&	50	&	0		&	8&7e-3	&	0&10	&	0&884	&&	4&5e-3	&	0&032	&	0&940	\\
specified	&	specified	&			&	1		&	8&0e-4	&	0&10	&	0&885	&&	2&5e-3	&	0&031	&	0&938	\\
	&		&			&	10	&	-0&16	&	0&15	&	0&806	&&	5&0e-3	&	0&048	&	0&983	\\
\cline{3-17}
	&		&	500	&	0	&	-2&4e-4	&	9&8e-3	&	0&919	&&	8&6e-4		&	3&3e-3	&	0&952	\\
	&		&	&	1			&	-3&2e-3	&	9&7e-3	&	0&924	&&	-9&3e-4	&	3&1e-3	&	0&953	\\
	&		&	&	10		&	-0&025	&	0&011	&	0&902	&&	-1&0e-3		&	4&5e-3	&	0&991	\\
\cline{3-17}
	&		&	5000	&	0	&	3&5e-4	&	9&9e-4	&	0&921	&&	-1&8e-4	&	3&1e-4	&	0&956	\\
	&		&	&	1				&	-2&7e-4	&	9&8e-4	&	0&922	&&	-1&4e-4	&	3&1e-4	&	0&950	\\
	&		&	&	10			&	-4&7e-3	&	1&0e-3	&	0&915	&&	-1&2e-4	&	4&5e-4	&	0&991	\\
\hline
Correctly	&	Mis-	&	50	&	0		&	0&025	&	0&10	&	0&934	&&	4&2e-3	&	0&030	&	0&954	\\
specified	&	specified	&			&	1		&	0&011	&	0&10	&	0&919	&&	1&8e-3	&	0&029	&	0&955	\\
	&		&			&	10	&	-0&18	&	0&28	&	0&933	&&	4&1e-3	&	0&030	&	0&952	\\
\cline{3-17}
	&		&	500	&	0	&	8&3e-3	&	9&9e-3	&	0&966	&&	-6&6e-4	&	3&0e-3	&	0&968	\\
	&		&	&	1			&	1&6e-3	&	9&8e-3	&	0&959	&&	8&0e-5		&	3&1e-3	&	0&965	\\
	&		&	&	10		&	-0&082	&	0&029	&	0&973	&&	-3&1e-4	&	2&9e-3	&	0&968	\\
\cline{3-17}
	&		&	5000	&	0	&	7&6e-3	&	1&1e-3	&	0&963	&&	2&9e-4		&	3&0e-4	&	0&967	\\
	&		&	&	1				&	1&8e-3	&	9&8e-4	&	0&964	&&	-1&8e-4	&	3&1e-4	&	0&967	\\
	&		&	&	10			&	-0&054	&	5&4e-3	&	0&916	&&	3&2e-4		&	2&9e-4	&	0&966	\\
\hline
\\
\end{tabular}
\end{center}
\end{table}

Results from this study are summarized in Table~\ref{t:one}, and additional simulation results for the overall effect under complete randomization are available in Table 3. We focus our discussion on the results in Table 2. Coverage probabilities in correctly-specified-model settings with interference are quite comparable to those without interference. This holds even under the moderate sample size of 50, with the exception of the direct effect when $\beta=10$. This exceptional setting has greater finite-sample bias, likely due to a combination of modest sample size and higher variance of $Y$ when $\beta=10$. Coverage is generally better for the overall effect parameter than for the direct effect. This is likely a result of the probability weights being more stable for the overall effect parameter, due to the OERS being closer to the actual propensity score than the interventions assigning all subjects to exposure and no exposure. Double robustness merely implies consistency, so correct coverage is not necessarily expected in the settings with model mis-specification, though we do still see fairly decent coverage in these settings. In practice, we recommend estimating the nuisance functions under more adaptive models using machine learning tools such as super learner, so as to avoid model mis-specification, and preserve valid inference.

There is a general trend of bias increasing with $\beta$, again, likely due to the increase in variance of $Y$. MSE remains relatively stable across $\beta$ values, indicating that the conditional variance given $\bar{A}$ is stable with respect to $\beta$, and that bias is small relative to conditional variance. Conditional variance is a more relevant metric than marginal variance, since our target parameter is adaptive via $\bar{A}$. The double robustness property of our estimator is evidenced by the decrease in bias and MSE with sample size in each setting.

\section{Conditions for Theorems 3 and 4}
Define the following: 
$\gamma_n\equiv\{W^n,A^n\}$; $\check{P}_{\gamma_n}$ to be the conditional distribution given $\gamma_n$; 
the stochastic processes
\[Z^Y_{\gamma_n,i}(g)\equiv D^c_{g^*_{\theta_n}}\left(\bar{Q}_{\theta_n,0},g\right)/\sqrt{n}=n^{-1/2}\frac{g^*_{\theta_n}(A_i\mid W_i)}{g(A_i\mid W_i)}\left\{Y_i-\bar{Q}_{\theta_n,0}(W_i,A_i)\right\}\]
for each $i$, indexed by the common semimetric space, $(\mathcal{F},\rho)$; the random semimetric
\[d^2_n(f,g)=\sum_{i=1}^n\left\{Z_{\gamma_n,i}(f)-Z_{\gamma_n,i}(g)\right\}^2;\]
$D^A_{g^*_{\theta_n}}(\bar{Q},g)\equiv \frac{g^*_{\theta_n}(A_i\mid W_i)}{g(A_i\mid W_i)}\bar{Q}(W_i,A_i)$; and the empirical process
\[Z_{\gamma_n,n}^A(\bar{Q})\equiv\sqrt{n}\left(\check{P}_{\gamma_n,0}-\tilde{P}_{\theta_n,0}\right)D^A_{g^*_{\theta_n}}(\bar{Q},g_{\theta_n,n}).\]
The conditions for Theorems 3 and 4 are:
\begin{enumerate}
\item Consistency of $\bar{Q}_{\theta_n,n}$ and $g_{\theta_n,n}$:
\[\max\left(\sup\limits_{\theta_n}\max\limits_{w\in w^n,a\in\{0,1\}}\lvert\bar{Q}_{\theta_n,n}-\bar{Q}_{\theta_n,0}\rvert,\sup\limits_{\theta_n}\max\limits_{w\in w^n,a\in\{0,1\}}\lvert g_{\theta_n,n}-g_{\theta_n,0}\rvert\right)\xrightarrow{p} 0\]
with the supremum taken over the support of $\{W^n,\bar{A}\}$, and

\begin{align*}
R_n&\equiv\frac{1}{n}\sum_{i=1}^n\sum_{a=0}^1\left\{\frac{g^*_{\theta_n}(a\mid W_i)}{g_{\theta_n,n}(a\mid W_i)}-\frac{g^*_{\theta_n}(a\mid W_i)}{g_{\theta_n,0}(a\mid W_i)}\right\}\left\{\bar{Q}_{\theta_n,0}(W_i,a)-\bar{Q}^*_{\theta_n,n}(W_i,a)\right\}g_{\theta_n,0}(a\mid W_i)\\
&=o_{p}(n^{-1/2}).
\end{align*}
\item Positivity: $g^*$ satisfies $\sup_{\theta_n,a\in\{0,1\}}\frac{g^*_{\theta_n}(a\mid W_i)}{g_{\theta_n,0}(a\mid W_i)} <\infty$ for $i\in\{1,\ldots,n\}$, with the supremum in $\theta_n$ taken over the support of $W^n$ and $\bar{A}$ in a neighborhood of $E_{P_0}(A)$.
\item Asymptotic equicontinuity of $Z^A_{\gamma_n,n}$: $Z^A_{\gamma_n,n}(\epsilon_n) = o_{p}(1)$ for any sequence $\epsilon_n$ converging to zero with respect to the supremum norm.
\item Measurability: The maps
\begin{align*}
(o_1,\ldots,o_n)&\mapsto\sup_{\rho(f,g)<\delta}\left\lvert\sum_{i=1}^n e_i\left\{Z^Y_{\gamma_n,i}(f)-Z^Y_{\gamma_n,i}(g)\right\}\right\rvert\\
(o_1,\ldots,o_n)&\mapsto\sup_{\rho(f,g)<\delta}\left\lvert\sum_{i=1}^n e_i\left\{Z^Y_{\gamma_n,i}(f)-Z^Y_{\gamma_n,i}(g)\right\}^2\right\rvert\\
\end{align*}
are measurable for every $\delta>0$, every vector $(e_1,\ldots,e_n)\in\{-1,0,1\}^n$, and every $n$.
\item Lindeburg condition: For every $\eta >0$,
\[\sum_{i=1}^nE\left\|Z^Y_{\gamma_n,i}\right\|^2_{\mathcal{F}}\left\{\left\|Z^Y_{\gamma_n,i}\right\|_{\mathcal{F}}>\eta\right\}\rightarrow 0,\]
where $\lVert X\rVert_{\mathcal{F}}\equiv \sup_{f\in\mathcal{F}}\lvert X(f)\rvert$.
\item Uniform asymptotic continuity of $\{Z^Y_{\gamma_n,i}\}_{i=1}^n$: For every $\delta_n\downarrow 0$, \[\sup_{\rho(f,g)<\delta_n}\sum_{i=1}^nE\left\{Z^Y_{\gamma_n,i}(f)-Z^Y_{\gamma_n,i}(g)\right\}^2\rightarrow 0\]
\item Entropy condition: For every $\delta_n\downarrow 0$,
\[\int_0^{\delta_n}\sqrt{\log \mathcal{N}(\varepsilon,\mathcal{F},d_n)}d\varepsilon\xrightarrow{P}0,\]
where $\mathcal{N}$ is the covering number of the set $\mathcal{F}$ for balls of radius $\varepsilon$ with respect to the semimetric $d_n$.
\end{enumerate}

\section{Proofs}
\begin{proof}(Theorem 1)
\begin{align*}
\psi^{g^*,\pi} &= \frac{1}{n}\sum_{i=1}^n E\left(Y_i^{g^*,\pi}\mid W^n\right)\\
&= \frac{1}{n}\sum_{i=1}^n \sum_{a=0}^1 E\left\{Y_i^{g^*,\pi}\mid W_i, A=a,k_n(\bar{A})=k_n(\pi)\right\}g^*_{\theta_n}(a\mid W_i)\\
&= \frac{1}{n}\sum_{i=1}^n \sum_{a=0}^1 E\{Y\mid W_i,a,k_n(\bar{A})=k_n(\pi)\}g^*_{\theta_n}(a\mid W_i)\\
&= \frac{1}{n}\sum_{i=1}^n \sum_{a=0}^1 \bar{Q}_0\{W_i,a,k_n(\pi)\}g^*_{\theta_n}(a\mid W_i),
\end{align*}
where the conditional expectation is well defined under Assumption 1, and
\begin{align*}
\psi^{g^*} &= \frac{1}{n}\sum_{i=1}^n E\left(Y_i^{g^*}\mid W^n\right)\\
&= \frac{1}{n}\sum_{i=1}^n \sum_{a^n} E\left\{Y_i^{g^*}\mid W_i, A=a_i,k_n(\bar{A})=k_n(\bar{a})\right\}g^*(a^n\mid W^n)\\
&= \frac{1}{n}\sum_{i=1}^n \sum_{a^n} E\left\{Y_i\mid W_i, A=a_i,k_n(\bar{A})=k_n(\bar{a})\right\}g^*(a^n\mid W^n)\\
&= \frac{1}{n}\sum_{i=1}^n \sum_{a^n} E\{Y\mid W_i,a,k_n(\bar{a})\}g^*(a^n\mid W^n),
\end{align*}
where the conditional expectation is well defined under Assumption 2.
\end{proof}

\begin{proof}(Theorem 2)
$\psi^{g^*,\bar{A}}=\Psi^{g^*,\bar{A}}(\bar{Q}_0)$ follows trivially from substituting $\bar{A}$ for $\pi$ in the causal identifying functional in Theorem 1. By Theorem 3, $\Psi^{g^*,\bar{A}}(\bar{Q}_0)$ can be estimated consistently. When $g^*$ is an ERS, we have
\begin{align*}
\psi^{g^*} &= \frac{1}{n}\sum_{i=1}^n \sum_{a^n} E\{Y\mid W_i,a,k_n(\bar{a})\}g^*(a^n\mid W^n)\\
 &= \frac{1}{n}\sum_{i=1}^n \sum_{a^n} E\{Y\mid W_i,a,k_n(\bar{A})\}g^*(a^n\mid W^n)\\
 &= \frac{1}{n}\sum_{i=1}^n \sum_{a=0}^1 E\{Y\mid W_i,a,k_n(\bar{A})\}g^*_{\theta_n}(a\mid W_i)\\
 &= \Psi^{g^*,\bar{A}}(\bar{Q}_0).
\end{align*}
\end{proof}

\begin{proof}(Theorem 3)

Let $\mathcal{M}_{\theta_n}$ denote the nonparametric model for $P_{\theta_n}(y^n,a^n)$, which leaves $Q_Y$ unrestricted, and $g_{\theta_n}$ constrained only by its relationship to $g$, which is unrestricted apart from the positivity condition. In $\mathcal{M}_{\theta_n}$, the TMLE is a substitution estimator based on estimates $\bar{Q}_{\theta_n,n}^{*}$ and $g_{\theta_n,n}$ of $\bar{Q}_{\theta_n,0}$ and $g_{\theta_n,0}$, respectively, that solve \[P_{\theta_n,n}D^c_{g^*_{\theta_n}}\left(Q_{\theta_n,n}^{*},g_{\theta_n,n}\right)=0.\]
Additionally,
\begin{align*}
&P_{\theta_n,0}D^c_{g^*_{\theta_n}}(\bar{Q}^*_{\theta_n,n},g_{\theta_n,n})\\
=& \frac{1}{n}\sum_{i=1}^n\sum_{a=0}^1\frac{g_{\theta_n}^*(a\mid W_i)}{g_{\theta_n,n}(a\mid W_i)}\left\{\int_yyq_{Y,0}(y\mid W_i,a,\theta_n)d\mu(y)-\bar{Q}^*_{\theta_n,n}(W_i,a)\right\}g_{\theta_n,0}(a\mid W_i)\\
=& \frac{1}{n}\sum_{i=1}^n\sum_{a=0}^1\frac{g^*_{\theta_n}(a\mid W_i)}{g_{\theta_n,n}(a\mid W_i)}\left\{\bar{Q}_{\theta_n,0}(W_i,a)-\bar{Q}^*_{\theta_n,n}(W_i,a)\right\}g_{\theta_n,0}(a\mid W_i)\\
=& \frac{1}{n}\sum_{i=1}^n\sum_{a=0}^1\left\{\bar{Q}_{\theta_n,0}(W_i,a)-\bar{Q}^*_{\theta_n,n}(W_i,a)\right\}g^*_{\theta_n}(a\mid W_i)\\
&+ \frac{1}{n}\sum_{i=1}^n\sum_{a=0}^1\left\{\frac{g^*_{\theta_n}(a\mid W_i)}{g_{\theta_n,n}(a\mid W_i)}-\frac{g^*_{\theta_n}(a\mid W_i)}{g_{\theta_n,0}(a\mid W_i)}\right\}\left\{\bar{Q}_{\theta_n,0}(W_i,a)-\bar{Q}^*_{\theta_n,n}(W_i,a)\right\}g_{\theta_n,0}(a\mid W_i)\\
=& \Psi(\bar{Q}_{\theta_n,0},Q_{W,n})-\Psi(\bar{Q}^*_{\theta_n,n},Q_{W,n})+R_n\\
=& \Psi(\bar{Q}_{\theta_n,0},Q_{W,n})-\Psi(\bar{Q}^*_{\theta_n,n},Q_{W,n})+o_p(n^{-1/2}).
\end{align*}
Then,
\begin{align*}
&\sqrt{n}\left\{\Psi(\bar{Q}^*_{\theta_n,n},Q_{W,n})-\Psi(\bar{Q}_{\theta_n,0},Q_{W,n})\right\}\\
=& \sqrt{n}\left(P_{\theta_n,n}-P_{\theta_n,0}\right)D^c_{g^*_{\theta_n}}(\bar{Q}^*_{\theta_n,n},g_{\theta_n,n})+o_p(1)\\
=& \sqrt{n}\left(P_{\theta_n,n}-\check{P}_{\gamma_n,0}\right)D^c_{g^*_{\theta_n}}(\bar{Q}_{\theta_n,n},g_{\theta_n,n}) + \sqrt{n}\left(\check{P}_{\gamma_n,0}-P_{\theta_n,0}\right)D^c_{g^*_{\theta_n}}(\bar{Q}_{\theta_n,n},g_{\theta_n,n})+o_p(1)\\
=& \sum_{i=1}^n Z^Y_{\gamma_n,i}(g_{\theta_n,n}) + \frac{1}{\sqrt{n}}\sum_{i=1}^n\frac{g^*_{\theta_n}(A_i\mid W_i)}{g_{\theta_n,n}(A_i\mid W_i)}\left\{\bar{Q}_{\theta_n,0}(W_i,A_i)-\bar{Q}_{\theta_n,n}(W_i,A_i)\right\}\\
&- \frac{1}{\sqrt{n}}\sum_{i=1}^n\sum_{a=0}^1\frac{g^*_{\theta_n}(a\mid W_i)}{g_{\theta_n,n}(a\mid W_i)}\left\{\bar{Q}_{\theta_n,0}(W_i,a)-\bar{Q}_{\theta_n,n}(W_i,a)\right\}g_{\theta_n,0}(a\mid W_i)+o_p(1)\\
=& \sum_{i=1}^n Z^Y_{\gamma_n,i}(g_{\theta_n,n}) + Z^A_{\gamma_n,n}(\bar{Q}_{\theta_n,0}-\bar{Q}_{\theta_n,n})+o_p(1)\\
=& \sum_{i=1}^n Z^Y_{\gamma_n,i}(g_{\theta_n,n}) + o_p(1),
\end{align*}
where the last equality follows by the asymptotic equicontinuity assumption on $Z^A_{\gamma_n,n}$.

An application of Theorem 2.11.1 in \cite{van1996weak} gives weak convergence of $\sum_{i=1}^nZ^Y_{\gamma_n,i}(\bar{Q}_{\theta_n,n},g_{\theta_n,n})$ to a Gaussian process with marginal variance $\sigma^2_Y \equiv\lim\limits_{n\rightarrow\infty}\sum_{i=1}^n\check{P}_{\gamma_n,0}Z^Y_{\gamma_n,i}\allowbreak (g_{\theta_n,0})^2$ at $g_{\theta_n,0}$, so \[\sum_{i=1}^nZ^Y_{\gamma_n,i}(g_{\theta_n,n})=\sum_{i=1}^nZ^Y_{\gamma_n,i}(g_{\theta_n,0})+o_p(1)\Rightarrow N(0,\sigma^2_Y).\]
Thus, we have convergence with respect to a sequence of conditional distributions, and 
\[\mathrm{Pr}\left\{\frac{\sum_{i=1}^nZ^Y_{\gamma_n,i}(g_{\theta_n,n})}{\sqrt{\sum_{i=1}^n\check{P}_{\gamma_n,0}Z^Y_{\gamma_n,i}(g_{\theta_n,0})^2}}\leq x\mathrel{\Bigg|} \gamma_n\right\}\rightarrow \Phi(x)\]
pointwise in $x$ for sequences $\gamma_n$. The function of the sequence $\{W_i,A_i\}_{i=1}^{\infty}$ mapping all points to one integrates to one under the true probability measure, and dominates the above probability over the entire support of $\{W^n,A^n\}$ for each $x$. Thus, by the dominated convergence theorem,
\[\int_{w^n}\sum_{a^n}\mathrm{Pr}\left\{\frac{\sum_{i=1}^nZ^Y_{\gamma_n,i}(g_{\theta_n,n})}{\sqrt{\sum_{i=1}^n\check{P}_{\gamma_n,0}Z^Y_{\gamma_n,i}(g_{\theta_n,0})^2}}\leq x\mathrel{\Bigg|} W^n=w^n,A^n=a^n\right\}\prod_{j=1}^ng_0(a_j\mid w_j)dQ_{W,0}(w_j)\]
converges pointwise to $\Phi(x)$, and the TMLE converges marginally to a normal distribution centered at $\Psi^{g^*,\bar{A}}(\bar{Q}_0)$ with variance $\sigma^2_Y$. 
\end{proof}

\begin{proof}(Theorem 4)
Define $\eta_n\equiv \bar{A}$ and the stochastic process $f_{W}(\eta_n)\allowbreak\equiv\allowbreak\sum_{a=0}^1\allowbreak\bar{Q}_{\eta_n,0}(W,\allowbreak a)\allowbreak g^*(a\mid W)$. Then
\begin{align*}
\sqrt{n}(\psi_n-\psi^{g^*}_p) &= \sqrt{n}(\psi_n-\psi^{g^*}_c)+\sqrt{n}(\psi^{g^*}_c-\psi^{g^*}_p)\\
&= \sqrt{n}(\psi_n-\psi^{g^*}_c)+\sqrt{n}\left\{\frac{1}{n}\sum_{i=1}^n\sum_{a=0}^1\bar{Q}_{\eta_n,0}(W_i,a)g^*(a\mid W_i)-\psi^{g^*}_p\right\}.
\end{align*}
By Theorem 3, the first term is asymptotically normal with asymptotic variance $\sigma_Y^2$. By taking a Taylor expansion of the second term around $\eta_n$ centered at $p\equiv\lim_{n\rightarrow\infty}\eta_n$, we have
\begin{align*}
\sqrt{n}(P_n-P_0)f_W(\eta_n) &= \sqrt{n}(P_n-P_0)f_W(p) + \sqrt{n}(P_n-P_0)\nabla_{\eta_n} f_W(\eta_n)\rvert_{\eta_n=p}(\eta_n-p)\\
&= \sqrt{n}(P_n-P_0)f_W(p) + o_p(1).
\end{align*}
By the central limit theorem, $\sqrt{n}(P_n-P_0)f_W(p)\Rightarrow N(0,\sigma_W^2)$.


Define $Z^W_i\equiv \left\{\sum_{a=0}^1\bar{Q}_{\eta_n,0}(W_i,a)g^*(a\mid W_i)-\psi^{g^*}_p\right\}/\sqrt{n}$. Then
\begin{align*}
&\mathrm{Pr}\left\{\frac{\sum_{i=1}^nZ^Y_{\gamma_n,i}(g_{\theta_n,n})}{\sqrt{\sum_{i=1}^n\check{P}_{\gamma_n,0}Z^Y_{\gamma_n,i}(g_{\theta_n,0})^2}}\leq x_y,\frac{\sum_{i=1}^n Z^W_i}{\sqrt{\sum_{i=1}^nP_0(Z^W_i)^2}}\leq x_w\right\}\\
=& \mathrm{Pr}\left\{\frac{\sum_{i=1}^nZ^Y_{\gamma_n,i}(g_{\theta_n,n})}{\sqrt{\sum_{i=1}^n\check{P}_{\gamma_n,0}Z^Y_{\gamma_n,i}(g_{\theta_n,0})^2}}\leq x_y\mathrel{\Bigg|}\frac{\sum_{i=1}^n Z^W_i}{\sqrt{\sum_{i=1}^nP_0(Z^W_i)^2}}\leq x_w\right\}\mathrm{Pr}\left\{\frac{\sum_{i=1}^n Z^W_i}{\sqrt{\sum_{i=1}^nP_0(Z^W_i)^2}}\leq x_w\right\}.
\end{align*}
The second factor converges pointwise to $\Phi(x_w)$ due to the central limit theorem. The first factor equals
\begin{align*}
& \int_{w^n}\sum_{a^n}\mathrm{Pr}\left\{\frac{\sum_{i=1}^nZ^Y_{\gamma_n,i}(g_{\theta_n,n})}{\sqrt{\sum_{i=1}^n\check{P}_{\gamma_n,0}Z^Y_{\gamma_n,i}(g_{\theta_n,0})^2}}\leq x_y\mathrel{\Bigg|}\frac{\sum_{i=1}^n Z^W_i}{\sqrt{\sum_{i=1}^nP_0(Z^W_i)^2}}\leq x_w,W^n=w^n,A^n=a^n\right\}\\
&\times dF_{W^n,A^n}\left\{w^n,a^n \mathrel{\Bigg|}\frac{\sum_{i=1}^n Z^W_i}{\sqrt{\sum_{i=1}^nP_0(Z^W_i)^2}}\leq x_w\right\}\\
=& \int_{w^n}\sum_{a^n}\mathrm{Pr}\left\{\frac{\sum_{i=1}^nZ^Y_{\gamma_n,i}(g_{\theta_n,n})}{\sqrt{\sum_{i=1}^n\check{P}_{\gamma_n,0}Z^Y_{\gamma_n,i}(g_{\theta_n,0})^2}}\leq x_y\mathrel{\Bigg|}\frac{\sum_{i=1}^n Z^W_i}{\sqrt{\sum_{i=1}^nP_0(Z^W_i)^2}}\leq x_w,W^n=w^n,A^n=a^n\right\}\\
&\times\frac{I\left[\frac{n^{-1/2}\sum_{i=1}^n \left\{\sum_{a=0}^1\bar{Q}_{\theta_n,0}(w_i,a)g^*(a\mid w_i)-\psi^p_{g^*}\right\}}{\sqrt{\sum_{i=1}^nP_0(Z^W_i)^2}}\leq x_w\right]}{\mathrm{Pr}\left\{\frac{\sum_{i=1}^n Z^W_i}{\sqrt{\sum_{i=1}^nP_0(Z^W_i)^2}}\leq x_w\right\}}\prod_{j=1}^ng_0(a_j\mid w_j)dQ_{W,0}(w_j)\\
=& \int_{w^n}\sum_{a^n}\mathrm{Pr}\left\{\frac{\sum_{i=1}^nZ^Y_{\gamma_n,i}(g_{\theta_n,n})}{\sqrt{\sum_{i=1}^n\check{P}_{\gamma_n,0}Z^Y_{\gamma_n,i}(g_{\theta_n,0})^2}}\leq x_y\mathrel{\Bigg|}W^n=w^n,A^n=a^n\right\}\\
&\times\frac{I\left[\frac{n^{-1/2}\sum_{i=1}^n \left\{\sum_{a=0}^1\bar{Q}_{\theta_n,0}(w_i,a)g^*_{\theta_n}(a\mid w_i)-\psi^{g^*}_p\right\}}{\sqrt{\sum_{i=1}^nP_0(Z^W_i)^2}}\leq x_w\right]}{\mathrm{Pr}\left\{\frac{\sum_{i=1}^n Z^W_i}{\sqrt{\sum_{i=1}^nP_0(Z^W_i)^2}}\leq x_w\right\}}\prod_{j=1}^ng_0(a_j\mid w_j)dQ_{W,0}(w_j)\\
\rightarrow & \Phi(x_y),
\end{align*}
since
\[\mathrm{Pr}\left\{\frac{\sum_{i=1}^nZ^Y_{\gamma_n,i}(g_{\theta_n,n})}{\sqrt{\sum_{i=1}^n\check{P}_{\gamma_n,0}Z^Y_{\gamma_n,i}(g_{\theta_n,0})^2}}\leq x_y\mathrel{\Bigg|}W^n=w^n,A^n=a^n\right\}\rightarrow\Phi(x_y)\]
pointwise in $x_y$ for all $\{w^n,a^n\}$. Thus,
\[\mathrm{Pr}\left\{\frac{\sum_{i=1}^nZ^Y_{\gamma_n,i}(g_{\theta_n,n})}{\sqrt{\sum_{i=1}^n\check{P}_{\gamma_n,0}Z^Y_{\gamma_n,i}(g_{\theta_n,0})^2}}\leq x_y,\frac{\sum_{i=1}^n Z^W_i}{\sqrt{\sum_{i=1}^nP_0(Z^W_i)^2}}\leq x_w\right\}\rightarrow\Phi(x_w)\Phi(x_y),\]
i.e.,
\[\left\{\frac{\sum_{i=1}^nZ^Y_{\gamma_n,i}(g_{\theta_n,n})}{\sqrt{\sum_{i=1}^n\check{P}_{\gamma_n,0}Z^Y_{\gamma_n,i}(g_{\theta_n,0})^2}},\frac{\sum_{i=1}^n Z^W_i}{\sqrt{\sum_{i=1}^nP_0(Z^W_i)^2}}\right\}\]
converges jointly to a multivariate normal distribution with identity covariance matrix. By Slutsky's theorem and the Cramer-Wold theorem, we have
\begin{align*}
\sqrt{n}(\psi_n-\psi_{g^*}^c)+\sqrt{n}\left\{\frac{1}{n}\sum_{i=1}^n\sum_{a=0}^1\bar{Q}_{\theta_n,0}(W_i,a)g^*_{\theta_n}(a\mid W_i)-\psi^{g^*}_p\right\}&=\sum_{i=1}^n\left\{Z^Y_{\gamma_n,i}(g_n)+Z_i^W\right\}\\
&\Rightarrow N(0,\sigma^2_Y+\sigma^2_W)
\end{align*}
\end{proof}

\begin{proof}(Theorem 5)
\begin{align*}
\psi^{g^*} =& \frac{1}{n}\sum_{i=1}^n\sum_{a^n}\bar{Q}_0\{W_i,a,k_n(\bar{a})\}\prod_{j=1}^ng^*_{\theta_n}(a_j\mid W_j)\\
=& \frac{1}{n}\sum_{i=1}^n\left[\sum_{a=0}^1\bar{Q}_0\{W_i,a,k_n(\bar{A})\}g^*_{\theta_n}(a\mid W_i)\right.\\
&\left.+ \sum_{a^n}\nabla_b \bar{Q}_0\{W_i,a_i,k_n(b)\}\vert_{b'}(\bar{a}-\bar{A})\prod_{j=1}^ng^*_{\theta_n}(a_j\mid W_j)\right]\\
=& \Psi^{g^*,\bar{A}}(\bar{Q}_0) + \frac{1}{n^2}\sum_{i=1}^n\sum_{a=0}^1\nabla_b\bar{Q}_0\{W_i,a,k_n(b)\}\vert_{b'}(a-\bar{A})g^*_{\theta_n}(a\mid W_i)\\
&+ \frac{1}{n}\sum_{i=1}^n\left[\sum_{a=0}^1\nabla_b\bar{Q}_0\left\{W_i,a,k_n(b)\right\}\vert_{b'}g^*_{\theta_n}(a\mid W_i)\frac{1}{n}\sum_{j:j\neq i}\sum_{a=0}^1(a-\bar{A})g^*_{\theta_n}(a\mid W_j)\right]\\
=& \Psi^{g^*,\bar{A}}(\bar{Q}_0) + \frac{1}{n^2}\sum_{i=1}^n\left(\sum_{a=0}^1\nabla_b\bar{Q}_0\{W_i,a,k_n(b)\}\vert_{b'}(a-\bar{A})g^*_{\theta_n}(a\mid W_i)\right.\\
&- \left.\left[\sum_{a=0}^1\nabla_b\bar{Q}_0\{W_i,a,k_n(b)\}\vert_{b'}g^*_{\theta_n}(a\mid W_i)\right]\left\{\sum_{a=0}^1(a-\bar{A})g^*_{\theta_n}(a\mid W_i)\right\}\right)\\
&+ \left[\frac{1}{n}\sum_{i=1}^n\sum_{a=0}^1\nabla_b\bar{Q}_0\left\{W_i,a,k_n(b)\right\}\vert_{b'}g^*_{\theta_n}(a\mid W_i)\right]\left[\frac{1}{n}\sum_{i=1}^n\sum_{a=0}^1ag^*_{\theta_n}(a\mid W_i) -\bar{A}\right]\\
=& \Psi^{g^*,\bar{A}}(\bar{Q}_0) + o_p(n^{-1/2}),
\end{align*}
where the last term is zero since $g^*_{\theta_n}$ is an AERS.
\end{proof}


\begin{table}
\begin{center}
\caption{Simulation results for overall effect parameter under complete randomization}
\centering
\begin{tabular}{c c l c r@{.}l r@{.}l r@{.}l}
\\
\hline\hline
$\bar{Q}$	&	$g$	&	$n$	&	$\beta$ &\multicolumn{2}{c}{Bias}&\multicolumn{2}{c}{MSE}&\multicolumn{2}{c}{CP}\\
\hline
Correctly	&	Correctly	&	50	&	0		&	0&0066	&	0&029	&	0&900\\
specified	&	specified	&	&	1		&	9&0e-5	&	0&028	&	0&903\\
	&		&	&	5		&	-0&011	&	0&028	&	0&905\\
	&		&	&	10	&	-0&066	&	0&043	&	0&831	\\
\cline{3-10}
	&		&	500	&	0		&	-0&00012	&	0&0027	&	0&949	\\
	&		&	&	1		&	0&00051		&	0&0027	&	0&946	\\
	&		&	&	5		&	0&00023		& 0&0026		&	0&948	\\
	&		&	&	10	&	-0&0080		&	0&0029	&	0&931	\\
\cline{3-10}
	&		&	5000	&	0		&	-0&00046	&	0&00025	&	0&959	\\
	&		&	&	1		&	-0&00024	&	0&00026	&	0&952	\\
	&		&	&	5		&	-0&00023	&	0&00027	&	0&949	\\
	&		&	&	10	&	-0&0013		&	0&00027	&	0&944	\\
\hline
Mis-	&	Correctly	&	50	&	0		&	0&0073	&	0&029	&	0&922	\\
specified	&	specified	&			&	1		&	0&0041	&	0&029	&	0&906	\\
	&		&			&	5		&	-0&013	&	0&028	&	0&946	\\
	&		&			&	10	&	-0&046	&	0&039	&	0&980	\\
\cline{3-10}
	&		&	500	&	0	&	0&0018		&	0&0027	&	0&959	\\
	&		&	&	1			&	-0&00045	&	0&0027	&	0&950	\\
	&		&	&	5			&	-0&0022		& 0&00267	&	0&972	\\
	&		&	&	10		&	-0&0084		&	0&0037	&	0&998	\\
\cline{3-10}
	&		&	5000	&	0	&	0&00046		&	0&00026	&	0&963	\\
	&		&	&	1				&	-0&00033	&	0&00026	&	0&958	\\
	&		&	&	5				&	-0&00017	&	0&00028	&	0&978	\\
	&		&	&	10			&	-0&0014		&	0&00034	&	0&999	\\
\hline
Correctly	&	Mis-	&	50	&	0		&	0&0036	&	0&025	&	0&895	\\
specified	&	specified	&			&	1		&	0&0021	&	0&025	&	0&895	\\
	&		&			&	5		&	-0&011	&	0&026	&	0&890	\\
	&		&			&	10	&	-0&072	&	0&037	&	0&816	\\
\cline{3-10}
	&		&	500	&	0	&	0&00065		&	0&0025	&	0&916	\\
	&		&	&	1			&	0&00057		&	0&0024	&	0&917	\\
	&		&	&	5			&	0&00012		& 0&0024		&	0&916	\\
	&		&	&	10		&	-0&011		&	0&0026	&	0&906	\\
\cline{3-10}
	&		&	5000	&	0	&	0&00015		&	0&00024	&	0&922	\\
	&		&	&	1				&	-0&00019	&	0&00024	&	0&922	\\
	&		&	&	5				&	-0&00013	&	0&00024	&	0&922	\\
	&		&	&	10			&	-0&0015		&	0&00026	&	0&911	\\
\hline
\\
\end{tabular}
\end{center}
\end{table}

\newpage

  \bibliographystyle{biom} 
 \bibliography{strat_interf}

\begin{thebibliography}{}

\bibitem[\protect\citeauthoryear{Abbring and Heckman}{Abbring and
  Heckman}{2007}]{abbring2007econometric}
Abbring, J.~H. and Heckman, J.~J. (2007).
\newblock Econometric evaluation of social programs, part {III}: Distributional
  treatment effects, dynamic treatment effects, dynamic discrete choice, and
  general equilibrium policy evaluation.
\newblock {\em Handbook of Econometrics} {\bf 6,} 5145--5303.

\bibitem[\protect\citeauthoryear{Aronow, Samii, et~al\mbox{.}}{Aronow
  et~al.}{2017}]{aronow2017estimating}
Aronow, P.~M., Samii, C., et~al. (2017).
\newblock Estimating average causal effects under general interference, with
  application to a social network experiment.
\newblock {\em The Annals of Applied Statistics} {\bf 11,} 1912--1947.

\bibitem[\protect\citeauthoryear{Balzer, Petersen, and van~der Laan}{Balzer
  et~al.}{2016}]{balzer2016targeted}
Balzer, L.~B., Petersen, M.~L., and van~der Laan, M.~J. (2016).
\newblock Targeted estimation and inference for the sample average treatment
  effect in trials with and without pair-matching.
\newblock {\em Statistics in Medicine} {\bf 35,} 3717--3732.

\bibitem[\protect\citeauthoryear{Basse and Feller}{Basse and
  Feller}{2017}]{basse2017analyzing}
Basse, G. and Feller, A. (2017).
\newblock Analyzing two-stage experiments in the presence of interference.
\newblock {\em Journal of the American Statistical Association} .

\bibitem[\protect\citeauthoryear{Cox}{Cox}{1958}]{cox1958planning}
Cox, D.~R. (1958).
\newblock {\em Planning of Experiments}.
\newblock New York: Wiley.

\bibitem[\protect\citeauthoryear{Gruber and van~der Laan}{Gruber and van~der
  Laan}{2010}]{gruber2010targeted}
Gruber, S. and van~der Laan, M.~J. (2010).
\newblock A targeted maximum likelihood estimator of a causal effect on a
  bounded continuous outcome.
\newblock {\em The International Journal of Biostatistics} {\bf 6,}.

\bibitem[\protect\citeauthoryear{Heckman, Lochner, and Taber}{Heckman
  et~al.}{1998}]{heckman1998general}
Heckman, J.~J., Lochner, L., and Taber, C. (1998).
\newblock General equilibrium treatment effects: A study of tuition policy.
\newblock Technical report, National Bureau of Economic Research.

\bibitem[\protect\citeauthoryear{Hudgens and Halloran}{Hudgens and
  Halloran}{2008}]{hudgens2008toward}
Hudgens, M.~G. and Halloran, M.~E. (2008).
\newblock Toward causal inference with interference.
\newblock {\em Journal of the American Statistical Association} {\bf 103,}
  832--842.

\bibitem[\protect\citeauthoryear{Imbens and Rubin}{Imbens and
  Rubin}{2015}]{imbens2015causal}
Imbens, G.~W. and Rubin, D.~B. (2015).
\newblock {\em Causal Inference in Statistics, Social, and Biomedical
  Sciences}.
\newblock Cambridge University Press.

\bibitem[\protect\citeauthoryear{Lassi, Cometto, Huicho, and Bhutta}{Lassi
  et~al.}{2013}]{lassi2013quality}
Lassi, Z.~S., Cometto, G., Huicho, L., and Bhutta, Z.~A. (2013).
\newblock Quality of care provided by mid-level health workers: Systematic
  review and meta-analysis.
\newblock {\em Bulletin of the World Health Organization} {\bf 91,} 824--833I.

\bibitem[\protect\citeauthoryear{Liu and Hudgens}{Liu and
  Hudgens}{2014}]{liu2014large}
Liu, L. and Hudgens, M.~G. (2014).
\newblock Large sample randomization inference of causal effects in the
  presence of interference.
\newblock {\em Journal of the American Statistical Association} {\bf 109,}
  288--301.

\bibitem[\protect\citeauthoryear{Ogburn, Sofrygin, Diaz, and van~der
  Laan}{Ogburn et~al.}{2017}]{ogburn2017causal}
Ogburn, E.~L., Sofrygin, O., Diaz, I., and van~der Laan, M.~J. (2017).
\newblock Causal inference for social network data.
\newblock {\em arXiv preprint arXiv:1705.08527} .

\bibitem[\protect\citeauthoryear{Robins}{Robins}{1998}]{robins1998marginal}
Robins, J.~M. (1998).
\newblock Marginal structural models.
\newblock {\em 1997 Proceedings of the American Statistical Association,
  Section on Bayesian Statistical Science} pages 1--10.

\bibitem[\protect\citeauthoryear{Rosenbaum}{Rosenbaum}{2007}]{rosenbaum2007interference}
Rosenbaum, P.~R. (2007).
\newblock Interference between units in randomized experiments.
\newblock {\em Journal of the American Statistical Association} {\bf 102,}
  191--200.

\bibitem[\protect\citeauthoryear{Rubin}{Rubin}{1980}]{rubin1980randomization}
Rubin, D.~B. (1980).
\newblock Randomization analysis of experimental data: The {F}isher
  randomization test comment.
\newblock {\em Journal of the American Statistical Association} {\bf 75,}
  591--593.

\bibitem[\protect\citeauthoryear{Sobel}{Sobel}{2006}]{sobel2006randomized}
Sobel, M.~E. (2006).
\newblock What do randomized studies of housing mobility demonstrate? {C}ausal
  inference in the face of interference.
\newblock {\em Journal of the American Statistical Association} {\bf 101,}
  1398--1407.

\bibitem[\protect\citeauthoryear{Sofrygin and van~der Laan}{Sofrygin and
  van~der Laan}{2015}]{sofrygin2015semi}
Sofrygin, O. and van~der Laan, M.~J. (2015).
\newblock Semi-parametric estimation and inference for the mean outcome of the
  single time-point intervention in a causally connected population.
\newblock {\em Journal of Causal Inference} .

\bibitem[\protect\citeauthoryear{Tchetgen~Tchetgen and
  VanderWeele}{Tchetgen~Tchetgen and VanderWeele}{2012}]{tchetgen2012causal}
Tchetgen~Tchetgen, E.~J. and VanderWeele, T.~J. (2012).
\newblock On causal inference in the presence of interference.
\newblock {\em Statistical Methods in Medical Research} {\bf 21,} 55--75.

\bibitem[\protect\citeauthoryear{Toulis and Kao}{Toulis and
  Kao}{2013}]{toulis2013estimation}
Toulis, P. and Kao, E. (2013).
\newblock Estimation of causal peer influence effects.
\newblock In {\em International Conference on Machine Learning}, pages
  1489--1497.

\bibitem[\protect\citeauthoryear{Tran, Yiannoutsos, Musick, Wools-Kaloustian,
  Siika, Kimaiyo, van~der Laan, and Petersen}{Tran
  et~al.}{2016}]{tran2016evaluating}
Tran, L., Yiannoutsos, C.~T., Musick, B.~S., Wools-Kaloustian, K.~K., Siika,
  A., Kimaiyo, S., van~der Laan, M.~J., and Petersen, M. (2016).
\newblock Evaluating the impact of a {HIV} low-risk express care task-shifting
  program: A case study of the targeted learning roadmap.
\newblock {\em Epidemiologic Methods} {\bf 5,} 69--91.

\bibitem[\protect\citeauthoryear{van~der Laan}{van~der
  Laan}{2014}]{van2014causal}
van~der Laan, M.~J. (2014).
\newblock Causal inference for a population of causally connected units.
\newblock {\em Journal of Causal Inference} {\bf 2,} 13--74.

\bibitem[\protect\citeauthoryear{van~der Vaart and Wellner}{van~der Vaart and
  Wellner}{1996}]{van1996weak}
van~der Vaart, A.~W. and Wellner, J.~A. (1996).
\newblock {\em Weak Convergence and Empirical Processes}.
\newblock Springer.

\bibitem[\protect\citeauthoryear{{World Health Organization}
  et~al\mbox{.}}{{World Health Organization} et~al.}{2012}]{world2012taking}
{World Health Organization} et~al. (2012).
\newblock Taking stock: Task shifting to tackle health worker shortages. 2007.
\newblock {\em Geneva: World Health Organization} .

\bibitem[\protect\citeauthoryear{Zheng and van~der Laan}{Zheng and van~der
  Laan}{2011}]{zheng2011cross}
Zheng, W. and van~der Laan, M.~J. (2011).
\newblock Cross-validated targeted minimum-loss-based estimation.
\newblock In {\em Targeted Learning}, pages 459--474. Springer.

\end{thebibliography}

\end{document}